\documentclass[aps,prb,twocolumn,superscriptaddress,showpacs]{revtex4-1}
\usepackage{amsmath,amssymb,amsfonts,amsthm}
\usepackage{braket}
\usepackage{graphicx}
\usepackage{bm}
\usepackage{bbm}
\usepackage{color}
\usepackage{booktabs,tabularx}
\usepackage{dcolumn}   
\usepackage{epstopdf}
\usepackage{bbold}
\usepackage[colorlinks=true,linkcolor=blue,urlcolor=blue,citecolor=blue]{hyperref}
\usepackage{commath}
\usepackage[caption=false]{subfig}
\usepackage{amsmath}

\begin{document}

 \newcommand{\breite}{1.0} 

\newtheorem{prop}{Proposition}
\newtheorem{cor}{Corollary}

\newcommand{\be}{\begin{equation}}
\newcommand{\ee}{\end{equation}}

\newcommand{\bea}{\begin{eqnarray}}
\newcommand{\eea}{\end{eqnarray}}

\newcommand{\Reals}{\mathbb{R}}     
\newcommand{\Com}{\mathbb{C}}       
\newcommand{\Nat}{\mathbb{N}}       

\newcommand{\id}{\mathbbm{1}}    

\newcommand{\Real}{\mathop{\mathrm{Re}}}
\newcommand{\Imag}{\mathop{\mathrm{Im}}}

\def\O{\mbox{$\mathcal{O}$}}   
\def\F{\mathcal{F}}			
\def\sgn{\text{sgn}}

\newcommand{\deo}{\ensuremath{\Delta_0}}
\newcommand{\dea}{\ensuremath{\Delta}}
\newcommand{\ak}{\ensuremath{a_k}}
\newcommand{\ad}{\ensuremath{a^{\dagger}_{-k}}}
\newcommand{\sx}{\ensuremath{\sigma_x}}
\newcommand{\sz}{\ensuremath{\sigma_z}}
\newcommand{\spl}{\ensuremath{\sigma_{+}}}
\newcommand{\smi}{\ensuremath{\sigma_{-}}}
\newcommand{\alk}{\ensuremath{\alpha_{k}}}
\newcommand{\bk}{\ensuremath{\beta_{k}}}
\newcommand{\ok}{\ensuremath{\omega_{k}}}
\newcommand{\vd}{\ensuremath{V^{\dagger}_1}}
\newcommand{\vi}{\ensuremath{V_1}}
\newcommand{\vo}{\ensuremath{V_o}}
\newcommand{\zc}{\ensuremath{\frac{E_z}{E}}}
\newcommand{\xc}{\ensuremath{\frac{\Delta}{E}}}
\newcommand{\xd}{\ensuremath{X^{\dagger}}}
\newcommand{\aok}{\ensuremath{\frac{\alk}{\ok}}}
\newcommand{\tpw}{\ensuremath{e^{i \ok s }}}
\newcommand{\tpe}{\ensuremath{e^{2iE s }}}
\newcommand{\tmw}{\ensuremath{e^{-i \ok s }}}
\newcommand{\tme}{\ensuremath{e^{-2iE s }}}
\newcommand{\epls}{\ensuremath{e^{F(s)}}}
\newcommand{\emis}{\ensuremath{e^{-F(s)}}}
\newcommand{\epl}{\ensuremath{e^{F(0)}}}
\newcommand{\emi}{\ensuremath{e^{F(0)}}}

\newcommand{\mkcomm}[1]{{\color{red}MK: #1}}

\newcommand{\lr}[1]{\left( #1 \right)}
\newcommand{\lrs}[1]{\left( #1 \right)^2}
\newcommand{\lrb}[1]{\left< #1\right>}
\newcommand{\nbt}{\ensuremath{\lr{ \lr{n_k + 1} \tmw + n_k \tpw  }}}

\newcommand{\om}{\ensuremath{\omega}}
\newcommand{\dw}{\ensuremath{\Delta_0}}
\newcommand{\wbp}{\ensuremath{\omega_0}}
\newcommand{\dv}{\ensuremath{\Delta_0}}
\newcommand{\vbp}{\ensuremath{\nu_0}}
\newcommand{\vplus}{\ensuremath{\nu_{+}}}
\newcommand{\vminus}{\ensuremath{\nu_{-}}}
\newcommand{\wplus}{\ensuremath{\omega_{+}}}
\newcommand{\wminus}{\ensuremath{\omega_{-}}}
\newcommand{\uv}[1]{\ensuremath{\mathbf{\hat{#1}}}} 
\newcommand{\avg}[1]{\left< #1 \right>} 

\let\underdot=\d 
\renewcommand{\d}[2]{\frac{d #1}{d #2}} 
\newcommand{\dd}[2]{\frac{d^2 #1}{d #2^2}} 

\newcommand{\pdd}[2]{\frac{\partial^2 #1}{\partial #2^2}} 
\newcommand{\pdc}[3]{\left( \frac{\partial #1}{\partial #2}
 \right)_{#3}} 
\newcommand{\matrixel}[3]{\left< #1 \vphantom{#2#3} \right|
 #2 \left| #3 \vphantom{#1#2} \right>} 
\newcommand{\grad}[1]{\gv{\nabla} #1} 
\let\divsymb=\div 
\renewcommand{\div}[1]{\gv{\nabla} \cdot #1} 
\newcommand{\curl}[1]{\gv{\nabla} \times #1} 
\let\baraccent=\= 
\definecolor{mjg}{rgb}{.08,.05,.8}
\newcommand{\mjg}[1]{{\color{mjg} #1}}


\title{Constructing quantum many-body scar Hamiltonians from Floquet automata}

\author{Pierre-Gabriel Rozon}
\affiliation{Physics Department, McGill University, Montr\'eal, Qu\'ebec H3A 2T8, Canada}
\email[]{}
\author{Michael J. Gullans}
\affiliation{Joint Center for Quantum Information and Computer Science, NIST/University of Maryland, College Park, Maryland, 20742 USA}
\author{Kartiek Agarwal}
\affiliation{Physics Department, McGill University, Montr\'eal, Qu\'ebec H3A 2T8, Canada}

\date{\today}
\begin{abstract}
We provide a systematic approach for constructing approximate quantum many-body scars (QMBS) starting from two-layer Floquet automaton circuits that exhibit trivial many-body revivals. We do so by applying successively more restrictions that force local gates of the automaton circuit to commute concomitantly more accurately when acting on select scar states. With these rules in place, an effective local, Floquet Hamiltonian is seen to capture dynamics of the automaton over a long prethermal window. We provide numerical evidence for such a picture and use our construction to derive several QMBS models, including the celebrated PXP model. 
\end{abstract}
\maketitle



\section{Introduction}
Understanding how thermalization arises from unitary evolution remains a fundamental challenge in the study of non-equilibrium quantum dynamics. The Eigenstate Thermalization Hypothesis~\cite{deutsch_quantum_1991,srednicki_chaos_1994} (ETH) postulates that eigenstates of many-body quantum systems themselves encode thermal correlations when viewed by a local observer. Although ETH has been numerically verified in a wide variety of quantum systems~\cite{rigolrelaxation2007,rigol_thermalization_2008,hyungwonstrongeth}, several important exceptions are known that challenge its associated dogma. The most prominent of these are integrable systems which occur in models with fine-tuned parameters~\cite{sutherland2004beautiful}, and many-body localized systems~\cite{BAA,PalHuse,nandkishore2016general,agarwal2017rare} where more robust local integrals of motion~\cite{oganesyanhuselocal} emerge due to strong disorder. These systems exhibit a lack of level repulsion at all energies, a hallmark of non-ergodicity, and have certain persistent quantum correlations~\cite{newtonscradle,Schreiber15,smith2016many}. 

More recently, an experiment in a chain of Rydberg atoms found dramatic revivals in many-body quantum correlations after apparent relaxation, only when the system is initialized in \emph{specific states}~\cite{bernien2017probing}. It is now understood that certain quantum systems can break ergodicity weakly~\cite{turner2018weak}, by only violating the ETH over a sub-extensive number of eigenstates. These systems have been dubbed quantum many-body scars~\cite{turner2018quantum} (QMBS), generalizing the phenomenon well known in the single-particle setting~\cite{HellerScar}. Since the initial findings, low entanglement eigenstates in the middle of the spectrum have been discovered in well known models~\cite{moudgalya2018exact,affleck2004rigorous} and a number of theoretical proposals for constructing new QMBS Hamiltonians have been put forth, with the aid of spectrum generating algebras~\cite{modugalyaSGA2020,markaklttowers2020,Choi2019,odeatunnelstotowers2020}, projective constructions~\cite{shiraishimori}, matrix product state representations~\cite{HoMPSScars}, among others; see Ref.~\cite{Moudgalya_2022} for a more exhaustive list of references. 

Crucially, these proposals yield Hamiltonians where the scar eigenstates are known \emph{exactly}. These scar eigenstates appear in group of degenerate eigenstates called \emph{towers}, where adjacent towers are separated in energy by the same amount $\Delta E$ . Low entanglement states can generally be constructed from these scar eigenstates and are seen to exhibit perfect revivals in correlations with a period $T \sim 1/\Delta E$ indefinitely. This is in contrast to the experimentally motivated PXP model~\cite{turner2018weak,turner2018quantum,serbyn2021quantum} which hosts approximate scar eigenstates and in which many body revivals decay over a long but finite duration. The corresponding scar towers are only approximately equidistant in energy, implying that low entanglement states obtained from a superposition of scarred eigenstates don't show perfectly regular revivals due to slow dephasing. Although weak perturbations may be added to exact QMBSs to obtain such a decay of revivals,
it remains a challenge to explain the existence of QMBS Hamiltonians such as the PXP model that have no small parameter, as well as uncover what sets the timescale for the decay of quantum revivals.


In this work, we illustrate general principles to derive (both exact and approximate) QMBS \emph{Hamiltonians} without any small parameters, starting from Floquet automaton \emph{(unitary) circuits}, and show how a timescale for the decay of revivals naturally emerges in this setting. Automatons have a long and rich history of study, arising from their intriguing dynamical properties in both the classical~\cite{wolfram1984cellular} and quantum settings~\cite{gopalakrishnan2018facilitated,klobas2021exact,wilkinson2020exact,iadecolafloquetqmbs}, and are often associated with systems with state space~\cite{fisher1961statistical,henley2010coulomb,chalker2017spin} or kinetic constraints~\cite{fredrickson1984kinetic,palmer1984models,jackle1991hierarchically,ritort2003glassy}. 

The Floquet automata considered in this work are unitary circuits that effect permutations of computational basis states on a chain of qubits (although more general automata can be adopted). For the automata considered, the Hilbert space is naturally fragmented into disjoint subspaces of computational basis states which are cycled through with successive applications of the automaton circuit. Thus, all computational basis states revive at fixed (but different) time intervals. It is natural to ask if these automata, which can be described as simple unitary circuits in the quantum setting, can be used to construct QMBS Hamiltonians which show similar revivals. We find that the answer is yes, and the principles uncovered can be used, for instance, to derive the PXP model, reveal timescales that govern the relaxation, 
and obtain new QMBS models that show revivals for arbitrarily chosen computational basis states. 

For concreteness, we focus on automata with a two-layer brickwork circuit, illustrated in Fig.~\ref{fig:illustrationRules}(a), which is composed of the elementary gate $U_0$ and whose Floquet unitary is given by $U_F = e^{-iA} e^{-iB}$, where $A, B$ are local Hamiltonians related by translation. Here, $A$ can be thought to be a sum of local, spatially disjoint Hamiltonians $A_j \equiv i \log U_{0,j}$ (to be made more precise later). A naive application of the Baker-Campbell-Hausdorff (BCH) formula to obtain a Hamiltonian from $U_F$ is bound to fail as higher order BCH terms blow up in amplitude quickly while growing more non-local. Instead, we ask when the local Hamiltonian, $H_{\text{eff}} = A+ B$, can reproduce dynamics generated by $U_F$ on a \emph{subspace} of chosen `orbit' states, by virtue of forcing higher-order BCH terms to remain small (or ideally vanish) in this subspace. In particular, defining $C_n (A,B)$ as the $n^{\text{th}}$ order term in the expansion, we formulate rules that strongly suppress $\norm{C_n (A,B) P_o}$, where $P_o$ is the projector onto the orbit subspace. Note that this bounds both $\norm{P_o C_n P_o}$, which governs the corrections to the dynamics within the subspace of chosen orbit states, and $\norm{(\mathbb{1}-P_o) C_n P_o}$, which governs the leakage from the orbit states into `generic' states. 

In fact, forcing all $C_n P_o$ terms to vanish identically ensures that the Hamiltonian $H_{\text{eff}} = A + B$ admits certain eigenstates that are eigenstates of both $A$ and $B$ separately\footnote{More precisely, the existence of a subspace $\mathbb{S}$ such that $[A^a,B^b]\mathbb{S} = 0$ for arbitrary integers $a,b$ is a sufficient and necessary condition for the existence of common eigenstates of $A$ and $B$}:---it is these select eigenstates, which if small in number, and possessing low entanglement, become the scarred eigenstates of the Hamiltonian $H_{\text{eff}}$. The latter is naturally the case if $A$ is composed of a set of spatially disjoint local Hamiltonians, for instance, as we assume. In fact, to derive QMBS Hamiltonians, a natural starting point may be to consider Hamiltonians $H = A + B$ and devise rules such that $A, B$ have a finite set of common, low-entanglement eigenstates. Importantly, the connection to an underlying automaton further guarantees that $e^{i A n} = \mathbb{1}$ for some integer $n$, and forces the eigenvalues of $A$ (and similarly $B$) to be equidistant in energy, another crucial property of QMBSs which leads to observable many body revivals. (In a separate work, it will be shown that all mid-spectrum excited states of the spin-1 AKLT model can be found by finding common eigenstates of appropriate partitions~\cite{moudgalya2018exact}.)

Beyond providing us with some principles to construct new QMBS Hamiltonians, the reference to automata also sheds light on the possible mechanism of decay of revivals in imperfect QMBSs. Two putative timescales emerge. First, the terms of the BCH expansion neglected in $H_{\text{eff}}$ give rise to leakage from ideal transition between orbit states as predicted by the automaton circuit; the corresponding timescale $\tau_l$ is governed by the inverse of $\norm{(1-P_o) C_n P_o}$ (for some finite $n$), and second, a prethermal timescale $\tau_p \sim e^{n_0}$ emerges that justifies the truncation of $H_{\text{eff}}$ to finite order---although the rules are designed to suppress BCH terms on orbit states, they eventually begin to grow at some higher order $n_0$. We find evidence of such phenomenology in the PXP model. 
In particular, there is an associated Floquet automaton~\cite{iadecolafloquetqmbs} which yields the PXP Hamiltonian upon truncation of the BCH series. We find that BCH terms initially \emph{decrease} with increasing order $n$, characteristic of the amplitude of terms in the Floquet-Magnus (FM) expansion~\cite{bukov2015universal} in the high frequency limit, with a period $T < 1$.  This behavior is suggestive of a prethermalization~\cite{mori2016rigorous,kuwahara2016floquet,abanin2017rigorous,agarwal2020dynamical} window $\tau_p \sim e^{1/T}$ wherein a truncated Hamiltonian can be justified. The parameter $T$ is an emergent timescale that comes from the suppression of commutators in our case and is not intrinsic to the two two-layer automaton which has a unit drive period. Next, we also find that adding higher order BCH terms to the PXP model improves revivals, up to the order above which the BCH series starts diverging again. Furthermore, these additional BCH terms correspond well with terms other authors have found using symmetry arguments in helping improve revivals in the PXP model~\cite{Choi2019,Khemani2019}. In this setup, the amplitude of these terms is fixed by the BCH expansion and not numerical optimization. 

This manuscript is organized as follows. In Sec.~\ref{sec:SetUp}, we detail the two-layer automata circuits we consider, with $U_F = U_A U_B$, discuss the fragmentation of the Hilbert space into sets of orbits, and the Floquet eigenstates of this system. We then discuss how we define the local Hamiltonians $A$, and $B$ from such automata. Sec.~\ref{sec:localrules} describes how we obtain a set of rules that can be used to generate scarred eigenstates in the effective Hamiltonian $H_{\text{eff}} = A + B$ and in particular embed certain (arbitrarily chosen) computational basis states in this scarred subspace. Sec.~\ref{sec:models} then describes a series of new models QMBS-A,B,C that we arrive at, using the methodology proposed, along with the PXP model. In Sec.~\ref{sec:NumericalEvidence}, we first provide evidence that the models show scar phenomenology and are non-integrable. The models QMBS-A,B,C exhibit successively stronger revivals (with QMBS-C exhibiting perfect revivals), in accordance with the fact that higher order BCH terms are more strongly suppressed in each successive model as per our construction. In Sec.~\ref{sec:BCHExpansionAndRevivals}, we discuss the amplitude of terms in the BCH expansion which connects the automaton to the Hamiltonian---for the PXP model, we find the amplitude of these terms show similar non-monotonic behavior expected in systems driven at high frequencies, indicating the possibility of a prethermalization window; adding more BCH terms to the PXP model also appears to improve revival strength and regularity. The evidence for such behavior is, however, limited in the other models we study. We end with Sec.~\ref{sec:DiscussionOutlook} where we discuss some questions that are raised by this approach and which need further analysis, besides summarizing our findings. 
\section{Underlying cellular automaton and associated Hamiltonian}\label{sec:SetUp}
\subsection{Physical setting}
\begin{figure}
    \centering
    \includegraphics[width=0.46\textwidth]{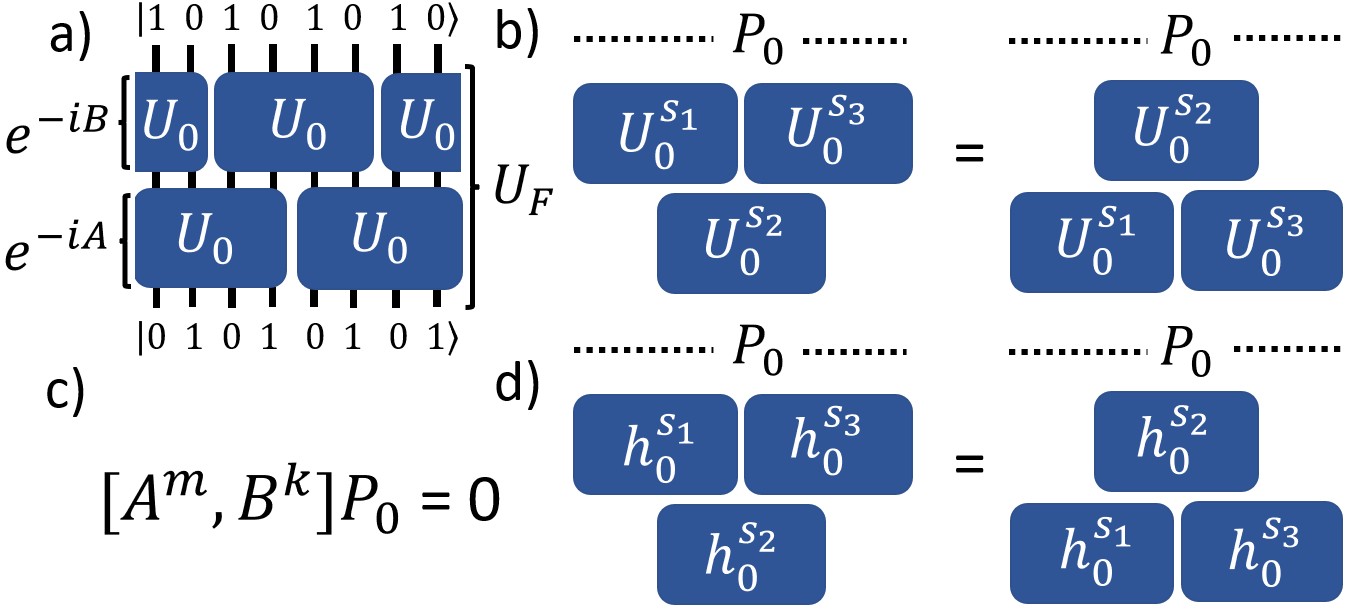}
    \caption[Structure of the quantum cellular automaton studied in this work, illustration of global rules and local rules of type I/II.]{a) A quantum cellular automaton (Floquet unitary $U_F$) that converts one N\'eel state to another; b) and d) Local commutation rules of Type I and II respectively enforced on the orbit subspace; c) Global rules.}
    \label{fig:illustrationRules}
\end{figure}

\begin{figure}
    \centering
    \includegraphics[width=0.30\textwidth]{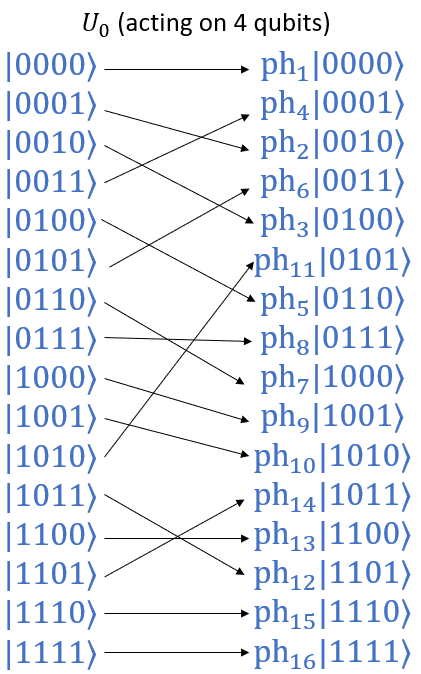}
    \caption[Illustration of the permutation nature of the gates $U_0$]{Example of a permutation gate $U_0$ acting on 4 adjacent qubits.}
    \label{fig:U0Permutation}
\end{figure}

\begin{figure}
\centering
   \includegraphics[width=0.4\textwidth]{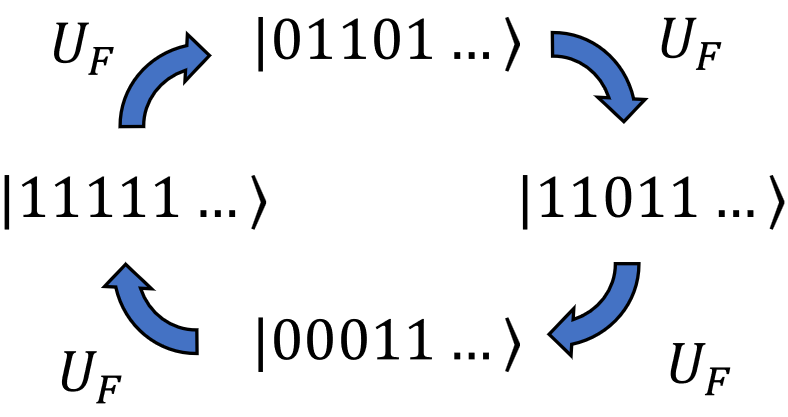}
   \caption{Example of a cycle of length 4 produced by the quantum cellular automaton $U_F$}\label{fig:cyclesIllustration}
\end{figure}

The quantum cellular automata considered in this work can be represented by a unitary circuit composed of two layers acting on a one-dimensional chain of $L$ qubits with periodic boundary conditions. The two layers combined are denoted by $U_F$, as shown in Fig.~\ref{fig:illustrationRules}(a).
Each layer is composed of local unitary gates $U_0$ that permute the computational basis states of the Hilbert space on which they act locally (the gates are chosen to have support on 4 qubits in this work) as well as multiplying them by a phase $\text{ph}_q$, see Fig, \ref{fig:U0Permutation}. Furthermore it is assumed that there exists an integer $n$ such that $U_{0}^n = \mathbb{1}$ which follows naturally from the permutation structure of the unitary $U_0$ provided the phases accrued also satisfy certain conditions; see Sec.~\ref{sec:UFEigenstates}.
As mentioned in the introduction, having $U_{0,j}^n = \mathbb{1}$, with $n$ finite is key to obtaining a scar subspace with equidistant eigenvalues embedded in an otherwise thermalizing spectrum. In the case of the $U_0$ considered in this work, $U_F$ itself is a permutation of the set of computational basis states that spawn the entire Hilbert space. This implies that $U_F$ can be decomposed into a set of disjoint cycles containing successive computational basis states obtained upon successive application of $U_F$ to a given state, see Fig.~\ref{fig:cyclesIllustration}. This fact can be used to solve exactly for the Floquet eigenstates of $U_F$, as discussed in Sec.~\ref{sec:UFEigenstates}.

The first layer of the circuit can be described as the exponential of a Hamiltonian $B$ such that $e^{-iB}$ yields the first layer of the circuit. Similarly, the second layer is associated with a Hamiltonian $A$, see Fig. ~\ref{fig:illustrationRules}(a). The exact definition of $A$ and $B$ is given in Sec.~\ref{sec:HFromU}. 

The sites on which the automata acts are labeled with the index $j \in \{1,2,3,4,...,L \}$. The local unitary gate $U_{0,j}$ is defined to act on the sites  $\{j,j+1,j+2,j+3\}$.
With this notation, the unitaries corresponding to the first and second layers are 
\begin{equation}
\begin{aligned}
    e^{-iB} = \prod_{j=1}^{L/4} U_{0,4j - 1},\quad
    e^{-iA} = \prod_{j=1}^{L/4} U_{0,4j - 3};
\end{aligned}
\end{equation} see Fig.~\ref{fig:illustrationRules} (a). 

\subsection{Eigenstates and eigenvalues of \texorpdfstring{$U_F$}{} }\label{sec:UFEigenstates}
The eigenstates of $U_F$ can be obtained from the cycles that the computational basis states undergo upon evolution by $U_F$. Indeed, suppose that $U_F$ produces a cycle of length $l$ given by the sequence of computational basis states $\ket{q} \rightarrow \ket{\sigma(q)} \rightarrow \ket{\sigma^2(q)} ... \ket{\sigma^{l-1}(q)} \rightarrow \ket{q}$ where $\ket{q}$ represents the $q^{\text{th}}$ computational basis state, and $\sigma^{n}(q)$ corresponds to the $n$ consecutive applications of the permutation $\sigma$ associated with $U_F$ on the state $\ket{q}$ ($U_F$ simply permutes the computational basis states up to a phase). Supposing that the unitary $U_F$ only has matrix elements 0 or 1 (no phase is acquired due to $U_F$), one directly observes that the quantum state $\ket{q} + \ket{\sigma(q)} + \ldots + \ket{\sigma^{l-1}(q)}$ is an eigenstate of the Floquet unitary with an eigenvalue of $1$. More generally, it is easy to show that states of the form 
\begin{equation}\label{eq:minimalEigenstates}
\ket{m,q_1} = \frac{1}{\sqrt{l}}\sum_{k = 0}^{l-1}e^{i\alpha_{k}}U_F^{k}\ket{q_1}
\end{equation}
with 
\begin{equation}
\begin{aligned}\label{eq:ConditionEigenstates}
    \beta = \frac{\Phi + 2\pi m}{l} \qquad \Phi = -i\log(\bra{q_1}U_F^l\ket{q_1})\\
    \alpha_{k} = -k\beta \qquad m \in \{0,1,...,l-1\}
\end{aligned}
\end{equation} 
form a complete orthonormal eigenbasis of $U_F$, where $\ket{q_1}$ in Eq.  ~\ref{eq:minimalEigenstates} is a computational basis state appearing in a given cycle of length $l$ and $e^{i\beta}$ is the eigenvalue of the state $\ket{m,q_1}$. For a given $\ket{q_1}$, distinct values of $m$ yield distinct eigenvalues which implies that the obtained states are mutually orthogonal. Eigenstates corresponding to different cycles are composed of different computational basis states, so they necessarily are orthogonal to each other. Thus, a complete orthonormal basis can be obtained by selecting a representative state $\ket{q_1}$ in each cycle, and the eigenvalue $m$. Note that if an integer $n$ such that $U_0^n = \mathbb{1}$ is to exist, it must be the case that \emph{all} the $\beta$ are integer fractions of $2\pi$ which is equivalent to requiring that all $\Phi$ associated with distinct $\ket{q_1}$ are an integer fraction of $2\pi$.

Eqs.~(\ref{eq:minimalEigenstates},\ref{eq:ConditionEigenstates}) show that Floquet eigenstates $\ket{q_1,m}$, with $m \in \{0,...,l-1\}$, when viewed as eigenstates of a corresponding Floquet Hamiltonian $H_F$ (such that $e^{-iH_Ft} = U_F$), correspond to eigenstates separated by a multiple of the constant energy difference $\Delta E = \frac{2\pi}{l}$. The computational basis states that appear in a given cycle of small length are thus ideal candidates as area-law entanglement states to embed in a physical model related to $U_F$. How this can be done is discussed in Sec.~\ref{sec:localrules}. First, however, we discuss how Hamiltonians $A$ and $B$ are precisely defined from the two-layer automata considered. 
 
\subsection{Local Hamiltonians from quantum cellular automata}\label{sec:HFromU}
We note that Hamiltonians $A$ and $B$ are not uniquely defined from $U_F$---there exist multiple Hamiltonians that that yield $U_F$ when exponentiated. Since a single layer of $U_F$ is composed of spatially decoupled unitaries $U_0$, we can also construct $A$ and $B$ from local Hamiltonians satisfying

\begin{equation}
h_{0,j} = i\log{U_{0,j}}.
\end{equation}

This equation alone does not uniquely specify $h_{0,j}$, but this ambiguity can be lifted by writing $U_{0,j}$ in terms of the orthonormal Floquet eigenstates obtained from Eq.~\ref{eq:minimalEigenstates} which yields
\begin{equation}
    U_{0,j} = \sum_{k=1}^{2^4}e^{i\beta_k}\ket{\beta_k}\bra{\beta_k}.
\end{equation}
$\ket{\beta_k}$ are the Floquet eigenstates of $U_{0,j}$ as defined in Eq. ~\ref{eq:minimalEigenstates}. $h_{0,j}$ is then defined as
\begin{equation}\label{eq:localLog}
h_{0,j} \equiv -\sum_{k=1}^{2^4}\tilde{\beta_k}\ket{\beta_k}\bra{\beta_k}
\end{equation}
where $\tilde{\beta}_k$ is $-i$ times the principal logarithm of $e^{i\beta_k}$, implying $\tilde{\beta}_k \in (-\pi,\pi]$. The Hamiltonian which we will force to support quantum scars is the strictly local Hamiltonian
\begin{equation}
H = A + B
\end{equation}
with $A = \sum_{j = 1}^{L/4}h_{0,4j - 3}$ and $B = \sum_{j = 1}^{L/4}h_{0,4j - 1}$. $A$ can thus be understood as the logarithm of the second layer of $U_F$ and $B$ as the logarithm of the first layer; see Fig.~\ref{fig:illustrationRules}(a). 

\subsection{Distinction between \texorpdfstring{$U_F$}{} and \texorpdfstring{$e^{-i(A + B)}$}{} }
So far nothing guarantees that the extracted Hamiltonian $A+B$ mimics the underlying quantum cellular automaton in any meaningful way. This is because the original Floquet unitary $U_{F} = e^{-iA}e^{-iB}$ and the time evolution operator associated with the $A + B$ Hamiltonian $e^{-i(A + B)}$ at $t = 1$ are not equal in general. The reason for this discrepancy can be understood once we interpret the automaton $U_F$ as the result of a periodic driving of the system. Indeed, successive applications of the Floquet unitary $U_F$ to a quantum state $\ket{\psi}$ is equivalent to the time evolution at even integer times resulting from the stroboscopic driving of the quantum system with the Hamiltonians $H = A$, $H = B$ for equal times.
The floquet unitary $U_F$ can alternatively be captured by a Floquet Hamiltonian $H_F$ such that $U_F = e^{-iH_F}$; $H_F$ is formally given by the Floquet-Magnus expansion. In particular, this expansion reduces to the well known Baker-Campbell-Hausdorff (BCH) expansion in the case of the driving described above. 

The first few terms of the BCH expansion are given by 
\begin{equation}\label{eq:BCHExpansion}
\begin{aligned}
H_{\text{F}} = A + B -\frac{i}{2}[A,B] -\\ \frac{1}{12}( [A,[A,B]] - [B,[A,B]]) + ...\quad.  
\end{aligned}
\end{equation}
The $i^{th}$ BCH term is denoted by $C_i$, where the $0^{th}$ order term is $A + B$. For instance,
\begin{equation}
\begin{aligned}
C_0 = A + B \quad C_1 = \frac{-i}{2}[A,B] \\ C_2 = -\frac{1}{12}( [A,[A,B]] - [B,[A,B]]), ...\quad. 
\end{aligned}
\end{equation}
Importantly, the amplitude of terms in this series quickly diverges, owing to the proliferation of the number of non-zero commutators of local terms. This implies that $H_{\text{F}}$ cannot generally be approximated by its first order term $A + B$ and thus $e^{-i(A + B)}\ket{\psi} \neq e^{-iA}e^{-iB}\ket{\psi}$ in general.  However, as we will show in the next sections, it is possible to obtain sets of rules which, if all or part of them are satisfied, ensure that some of the subspaces associated with cycles of $U_F$ are preserved or approximately preserved by $H = A+B$. One useful set of local rules can be obtained by realizing that the local Hamiltonians $h_{0,j}$ assume a special decomposition in terms of powers of $U_{0,j}$ as we discuss next.

\section{Rules that guarantee the presence of quantum scars}\label{sec:localrules}
\subsection{Writing H as a linear superposition of powers of simple unitary gates}\label{sec:HasUPowers}
The local unitary gates considered in this work are chosen such that there exists an integer $n$ for which $U_{0,j}^n = \mathbb{1}$. Provided $U_{0,j}^n = \mathbb{1}$, along with the definition of $h_{0,j}$ specified in Eq.~\ref{eq:localLog}, one can show that
\begin{equation}\label{eq:logUexpansion}
    h_{0,j} = i\log{U_{0,j}} = \sum_{k = 0}^{m-1}c_kU_{0,j}^k
\end{equation}
for some set of coefficients $c_k$.
An exact recipe for obtaining the coefficients $c_k$ is given in App.~\ref{app:SolvingForC}; we note here that in Eq.~(\ref{eq:logUexpansion}), the integer $m \le n$ (where $U_{0,j}^n = \mathbb{1}$). In other words, it is possible that not all powers of $U_{0,j}$ up to $n$ are required to construct $h_{0,j}$. This is the case for the PXP model for which $U_{0,j}^4 = \mathbb{1}$, but $\mathbb{1}, U_{0,j}, U^2_{0,j}$ are sufficient to obtain $h_{0,j} = \text{PXP}_j$; see Tab.~\ref{tab:PXPTabSpin} for a definition of $U_{0,j}$ and $h_{0,j}$ in the PXP model.

\subsection{Global rules}
Eq.~(\ref{eq:logUexpansion}) can be leveraged to construct a set of rules that will ensure that some chosen area law entanglement states are common eigenstates of $A$ and $B$. Indeed, provided a decomposition of $h_{0,j}$ in terms of powers of $U_{0,i}$, one can rewrite $A$ and $B$ as
\begin{equation}
    A = \sum_{j = 1}^{L/4}\sum_ {k = 1}^{n}c_kU_{0,4j -3 }^k,
\end{equation}
\begin{equation}
    B = \sum_{j = 1}^{L/4}\sum_{k = 1}^{n}c_kU_{0,4j -1 }^k. 
\end{equation}
Next, consider the subspace spawned by a specific cycle of $U_F$ that has length $l$ and define the projector $P_o$ to be the projector onto the computational basis states that compose the cycle
\begin{equation}
    P_0 = \sum_{k=1}^l\ket{\sigma^k(q)}\bra{\sigma^k(q)}.
\end{equation}
The states that appear in this cycle are the area-law entanglement states chosen here to be embedded as a linear superposition of common eigenstates of $A$ and $B$. A sufficient condition to embed the subspace spawned by $P_0$ is to enforce that $e^{-i(A + B)}P_0$ yields the same result as $e^{-iA}e^{-iB}P_0$. For this to be true, it is sufficient to require that
\begin{equation}\label{OrbitCondition}
    [A^a,B^b]P_o = 0 \quad \forall a,b
\end{equation}
where $a,b$ are positive integers. This set of rules is a necessary and sufficient condition for the existence of a set of common eigenstates \cite{SHEMESH198411} of $A$ and $B$ denoted here by $\mathbb{S}$ which will spawn the computational basis states that appear in $P_0$. Such rules are dubbed global rules, see Fig.~\ref{fig:illustrationRules}(c). Satisfaction of all such global rules ensures QMBS phenomenology since the dynamical evolution of the computational basis states that appear in $P_0$ undergo a periodic cycle in accordance with the dynamical evolution prescribed by the underlying Floquet automaton instead of quickly thermalizing. Furthermore, provided the dimension of the common eigenstate subspace $\mathbb{S}$ grows at most polynomially with system size, then the common eigenstates of $A$ and $B$ will necessarily have low entanglement since linear combinations of such states must spawn the low entanglement states that appear in $P_0$. As a consequence, the common eigenstates of $A$ and $B$ appear as scar eigenstates of $H = A + B$ and form scar towers.

Since $A$ and $B$ are sums of spatially decoupled unitary gates, powers of $A$ and $B$ are given by
\begin{equation}
\begin{aligned}
     A^a = \left(\sum_{j=1}^{L/4}\sum_{k=1}^{n}c_kU_{0,4j-3}^k\right)^a \\ B^b = \left(\sum_{j=1}^{L/4}\sum_{k=1}^{n}c_kU_{0,4j-1}^k\right)^b 
\end{aligned}
\end{equation} and generic terms in $A^aB^bP_0$ take the form
\begin{equation}
\prod_{j = 1}^{L/4} U_{0,4j - 3}^{\alpha_{4j-3}}\prod_{j = 1}^{L/4} U_{0,4j - 1}^{\alpha_{4j-1}}P_o
\end{equation} up to a multiplicative constant, for some set of positive integers $\alpha_j$ including $0$.
Thus, in order to satisfy the identity $[B^{b},A^{a}]P_o = 0 $ for arbitrary integers $a$ and $b$, it is sufficient to require that the expression
\begin{equation}\label{FundamentalRelation}
\prod_{j = 1}^{L/4} U_{0,4j - 3}^{\alpha_{4j-3}}\prod_{j = 1}^{L/4} U_{0,4j - 1}^{\alpha_{4j-1}}P_o =\prod_{j = 1}^{L/4} U_{0,4j - 1}^{\alpha_{4j-1}}\prod_{j = 1}^{L/4} U_{0,4j - 3}^{\alpha_{4j-3}}P_o  
\end{equation}
is satisfied for all possible set of $\alpha_j$. Note that a distinct condition can be obtained by considering the alternate representation
\begin{equation}
\begin{aligned}
     A^a = \left(\sum_{j=1}^{L/4}h_{0,4j-3}\right)^a \quad B^a = \left(\sum_{j=1}^{L/4}h_{0,4j-1}\right)^b 
\end{aligned}
\end{equation}
which leads to the condition
\begin{equation}\label{FundamentalHamiltonianRelation}
\prod_{j = 1}^{L/4} h_{0,4j - 3}^{\alpha_{4j-3}}\prod_{j = 1}^{L/4} h_{0,4j - 1}^{\alpha_{4j-1}}P_o =\prod_{j = 1}^{L/4} h_{0,4j - 1}^{\alpha_{4j-1}}\prod_{j = 1}^{L/4} h_{0,4j - 3}^{\alpha_{4j-3}}P_o.
\end{equation}
As will be discussed next, conditions (\ref{FundamentalRelation}) and (\ref{FundamentalHamiltonianRelation}) lead to distinct sets of local rules, dubbed rules of type I and II respectively.
\subsection{Local rules of type I}
\begin{figure}
   \includegraphics[width=0.46\textwidth]{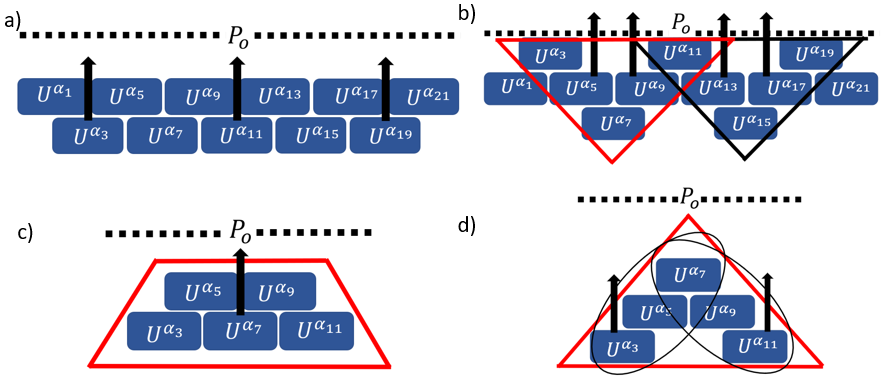}
   \caption[Illustration of the proof that local rules are sufficient for ensuring the embedding of common eigenstates.]{ a) Half the gates from the second layer are sent forward producing a circuit with three layers. b) Within each triangle (focusing on the red one), gates from the second layer are sent forward. c) Within the resulting configuration, the middle gate of the second layer is sent to the first layer d) The side gates are sent from the third layer to the first layer (Note that this is a 2 step operation for each side gate). The resulting arrangement of gates shows that by making use of the commutation rules, it is possible to send gate $U_0^{\alpha_7}$ from the third layer to the first layer in b). Repeating this procedure on each triangle proves that satisfying the local unitary rules is sufficient to ensure that $[A^a,B^b]P_0 = 0$ for arbitrary integers $a$ and $b$ 
}\label{FigureProof}
\end{figure}

Conditions (\ref{FundamentalRelation}) and (\ref{FundamentalHamiltonianRelation}) can be further reduced to simple local rules that only involve a small set of unitary gates. The set of local rules associated with condition (\ref{FundamentalRelation}) is given by

\begin{equation}\label{eq:typeIrules}
\begin{aligned}
    U_{0,j}^{s_1}U_{0,j+4}^{s_3}U_{0,j+2}^{s_{2}}\ket{\sigma^k(q)}  = U_{0,j+2}^{s_{2}}U_{0,j}^{s_1}U_{0,j+4}^{s_3}\ket{\sigma^k(q)} \\ \forall s_i \in \{0,1,2,...,n-1 \}, \quad \forall j \in \{1,3,5,...,L-1\}\\ U_{0,j}^n = \mathbb{1} \quad \forall k \in \{1,2,...,l\}
\end{aligned}
\end{equation}
where $\ket{\sigma^k(q)}$ are the states that appear in $P_0$. These rules are denoted rules of type I [see Fig.~\ref{fig:illustrationRules}(b)] and a graphical proof that they indeed ensure that Eq.~(\ref{FundamentalRelation}) is satisfied is provided in Fig.~\ref{FigureProof}. A remarkable property of type I rules is that they are finite and independent of the system size if the states $\ket{\sigma^k(q)}$ are translationally invariant. More precisely, given the smallest integer $m$ such that $S^{2m}\ket{\sigma^k(q)} = \ket{\sigma^k(q)}$ where $S$ is the operator translating all sites by one to the right, then the total number of sites $j$ one needs to check for the rules associated with the state $\ket{\sigma^k(q)}$ is reduced to $j \in \{1,3,...,2m-1\}$.

\subsection{Local rules of type II}
If condition~(\ref{FundamentalHamiltonianRelation}) is considered instead of condition~(\ref{FundamentalRelation}), one obtains a different set of local rules given by
\begin{equation}
\begin{aligned}
    h_{0,j}^{s_1}h_{0,j+4}^{s_3}h_{0,j+2}^{s_{2}}\ket{\sigma^k(q)}  = h_{0,j+2}^{s_{2}}h_{0,j}^{s_1}h_{0,j+4}^{s_3}\ket{\sigma^k(q)} \\ \forall s_i \in \{0,1,2,...,n-1 \}\quad \forall j \in \{1,3,5,...,L-1\} \\
    U_{0,j}^n = \mathbb{1} \quad \forall k \in \{1,2,...,l\}
    \label{eq:typeIIrules}
\end{aligned}
\end{equation}
which are denoted rules of type II, see Fig.~\ref{fig:illustrationRules}(d). A key distinction with rules of type II is that nothing ensures the existence of an integer $n$ such that $h_{0,j}^n = \mathbb{1}$. However, it is easy to see from the decomposition~(\ref{eq:logUexpansion}) that $h_{0,j}^n$ can always be written as a linear superposition of smaller powers of $h_{0,j}$; this restricts $s_i$ to be less than $n$; see App.~\ref{app:H0Closing} for more details. As discussed in Sec.~\ref{sec:HasUPowers}, there is a possibility that not all powers of $U_{0,j}$ up to $n$ are actually required to build $h_{0,j}$. This is also the case when considering a decomposition of $h_{0,j}$ in terms of smaller powers of itself. Indeed, there can exist an integer $m$ smaller then $n$ such that $h_{0,j}^m$ can be written as a linear superposition of smaller powers of $h_{0,j}$ which can further reduce the set of integers $s_i$ one actually needs to check. For instance, this is true in the PXP model for which $h_{0,j}^3 = \frac{\pi^2 }{4}h_{0,j}$. See Tab.~\ref{tab:PXPTabSpin} for the definition of the $h_{0,j}$ associated with the PXP model.

Another key distinction between rules of type I and rules of type II is that whenever a rule of type I is broken, BCH terms at all orders become non-vanishing. While since the BCH expansion is organised in terms of commutators of $h_{0,j}$, higher powers of $h_{0,j}$ in commutators only emerge at higher order in the BCH expansion. Thus, satisfying lower powers of the type II rules may be important in enforcing prethermal behavior and stabilising scar phenomenology in the truncated Hamiltonian (although there is no distinction between the two set of rules when all of them are satisfied). 

A final reason to consider type II rules is that one could in principle completely ditch any reference to automata and try to find Hamiltonians which satisfy these local rules to yield common eigenstates with low entanglement---the real purpose of the connection to an underlying automaton is to ensure scar phenomenology and to restrict the search for $h_{0,j}$ to Hamiltonians which yield a finite set of distinct operators $h^i_{0,j}$ with $i \in \{0,...,n-1\}$ 

\section{Building models that satisfy local rules}\label{sec:models}
It was shown in Sec.~\ref{sec:localrules} that satisfying all local rules is sufficient to ensure the protection of the subspace spawned by the states that appear in $P_0$. 

We note that the rules rely on two choices: i) the unitary $U_0$ which is determined, in the case we consider, by the permutation it generates over computational basis states, along with the phases accrued, and ii) the set of computational basis states $\ket{\sigma^k (q)}$, $k = 1,...,l$, we choose to embed in the putative scar subspace, the projector to which is given by $P_0$. Now, given the above structure, we note that the rules of type I, given in Eq.~(\ref{eq:typeIrules}), are either exactly satisfied (for a given choice of $s_1,s_2,s_3$ and $\ket{\sigma^k (q)}$, or the left and right hand side of Eq.~(\ref{eq:typeIrules}) produce entirely different computational basis states and/or phases. Thus, we can simply count the number of rules that are satisfied. The situation is trickier for the set of local rules given in Eq.~(\ref{eq:typeIIrules}), in that the local Hamiltonians $h_{0,j}$ will generically produce entangled states upon acting on computational basis states in $P_0$, and it may be useful to quantify the violation of the rules using a suitable inner product between the left and right hand sides of Eq.~(\ref{eq:typeIIrules}). For simplicity, for a search of model Hamiltonians with scar subspaces which we perform next, we restrict ourselves to rules of type I and simply enumerate the number of rules (out of a maximum determined by enumerating the allowed values of $s_1,s_2,s_3, k$). 

\subsection{Explicit model search}
There is a total of $16!$ permutations of the set of computational basis states that spawn the $4$ qubits Hilbert space on which $U_{0,j}$ acts and if phase is allowed, the space of possibilities is effectively infinite. The size of the search space makes it prohibitively hard to study exhaustively. To remedy this problem, we choose to restrict  $U_0$ to act trivially on the rightmost qubit while also preventing phase from being acquired. This produces a set of $8!$ possible unitary gates which can be studied exhaustively. The search space was further reduced by considering unitary gates such that $U_{0,j}^6 = \mathbb{1}$. The chosen subspace to protect is given by the two Néel states $\ket{q} = \ket{1,0,1,...}$, $\ket{\sigma(q)} = \ket{0,1,0,...}$, such that $U_F\ket{q} = \ket{\sigma(q)}, U_F \ket{\sigma(q)} = \ket{q}$. This constrained search results in three models presented in Tab.~\ref{tab:ModelsTab} which satisfy 70/350, 246/350 and 350/350 of the applicable type I rules, respectively. The unitary gates are represented in Tab.~\ref{tab:ModelsTab} by a permutation and a phase map (in this case trivial) which are defined in App.~\ref{app:PermPhaseDef}. The total number of relevant rules for each model is discussed in App.~\ref{app:TotNumberRules}
\subsection{PXP model}
Outside of this search, the PXP model is also studied in association with an underlying automaton. The circuit geometry is different due to the fact that the PXP model has a unit cell composed of one qubit, i.e $U_{F} = \prod_{j}^{L/2}U_{0,2j-1}\prod_{j}^{L/2}U_{0,2j}$ and $U_0$ in this case is a Toffoli gate which acts on three qubits. Note also that adjacent gates $U_{0,j}, U_{0,j+2}$ commute in the PXP model, so the first and second layer can be seen as a product of decoupled gates and the formalism developed in Sec.~\ref{sec:localrules} applies. Finally, for this model, the protected cycle is composed of three states instead of two and given by $\ket{q} = \ket{1,1,1,1,...}, \ket{\sigma(q)} = \ket{0,1,0,1,...}, \ket{\sigma^2(q)} = \ket{1,0,1,0...}$.

\begin{table}
\begin{center}
\begin{tabular}{||p{2cm}| p{6cm} ||}
 \hline\hline
 QMBS-A &  \\ [0.75ex] 
 \hline\hline
 Permutation & $\begin{aligned}((3, 13, 11, 7, 9, 5), \\(4, 14, 12, 8, 10, 6))\end{aligned}$ \\
 \hline
 Phase & (1,1,1,1,1,1,1,1,1,1,1,1,1,1,1,1)  \\
 \hline
 $h_{0,j}$ decomposition & $\begin{aligned}(\frac{\pi}{6} + i\frac{\pi}{2\sqrt{3}})U_{0,j} + (-\frac{\pi}{6} - i\frac{\pi}{6\sqrt{3}})U_{0,j}^2 +\\ \frac{\pi}{12}U_{0,j}^{3} - \frac{\pi}{12}U_{0,j}^0 + \text{h.c}\end{aligned}$  \\
 \hline
 $\begin{aligned}U_{0,j}^n = I\end{aligned}$ & n = 6 \\
 \hline
 Orbit & $\ket{q} = \ket{1,0,1,0,...}, \ket{\sigma(q)} = \ket{0,1,0,1,...}$\\ 
 \hline
 Rule ratio type I & $ 70/350$ \\
 \hline
 \hline
  QMBS-B &  \\ 
 \hline\hline
 
 Permuation & $\begin{aligned}((1, 15), (2, 16), (3, 9, 5), \\(4, 10, 6), (7, 13, 11),(8, 14, 12)) \end{aligned}$\\ 
 \hline
 Phase & (1,1,1,1,1,1,1,1,1,1,1,1,1,1,1,1)  \\
 \hline
 $h_{0,j}$ decomposition & $\begin{aligned}(\frac{\pi}{6} + i\frac{\pi}{2\sqrt{3}})U_{0,j} + (-\frac{\pi}{6} - i\frac{\pi}{6\sqrt{3}})U_{0,j}^2\\ + \frac{\pi}{12}U_{0,j}^{3} - \frac{\pi}{12}U_{0,j}^0 + \text{h.c}\end{aligned}$  \\
 \hline
 $U_{0,j}^n = \mathbb{1}$ & n = 6 \\
 \hline
 Orbit & $\ket{q} = \ket{1,0,1,0,...}, \ket{\sigma(q)} = \ket{0,1,0,1,...}$\\ [1ex] 
 \hline
 Rule ratio type I & 246/350 \\
 \hline
 \hline
QMBS-C & \\ 
\hline\hline
Permuation & $\begin{aligned}((3, 5), (4, 6), (7, 15, 9),\\ (8, 16, 10), (11, 13), (12, 14))\end{aligned}$\\ 
\hline
Phase & (1,1,1,1,1,1,1,1,1,1,1,1,1,1,1,1)  \\
\hline
$h_{0,j}$ decomposition & $\begin{aligned}(\frac{\pi}{6} + i\frac{\pi}{2\sqrt{3}})U_{0,j} + (-\frac{\pi}{6} - i\frac{\pi}{6\sqrt{3}})U_{0,j}^2 +\\ \frac{\pi}{12}U_{0,j}^{3} - \frac{\pi}{12}U_{0,j}^0 +\text{h.c}\end{aligned}$  \\
\hline

$U_{0,j}^n = \mathbb{1}$ & n = 6 \\
\hline
Orbit & $\ket{q} = \ket{1,0,1,0,...}, \ket{\sigma(q)} = \ket{0,1,0,1,...}$\\ [1ex] 
\hline
Rule ratio type I & 350/350 \\
\hline
\hline
PXP &  \\ [0.5ex] 
 \hline\hline
 Permuation & $((11, 15), (12, 16))$ \\ 
 \hline
 Phase & (1,1,1,1,1,1,1,1,1,1,i,i,1,1,i,i)  \\
 \hline
 $h_{0,j}$ decomposition & $(\frac{\pi}{4} + i\frac{\pi}{4})U_{0,j} -\frac{\pi}{8}U_{0,j}^2 -  \frac{\pi}{8}I +$ h.c \\
 \hline
 $U_0^n = \mathbb{1}$ & n = 4 \\
 \hline
 Orbit & $\begin{aligned}\ket{q} = \ket{1,1,1,1,...}, \ket{\sigma(q)} = \ket{0,1,0,1,...},\\ \ket{\sigma^2(q)} = \ket{1,0,1,0,...}\end{aligned}$\\ 
 \hline
 Rule ratio type II & 38/48 \\[1ex]
 \hline
\end{tabular}
\end{center}
\caption{Characteristics of the models}
\label{tab:ModelsTab}
\end{table}

\section{Numerical signature of quantum scars}\label{sec:NumericalEvidence}
\subsection{Revival strength and signs of quantum scarring}
\begin{figure}
    \centering
    \includegraphics[width=0.46\textwidth]{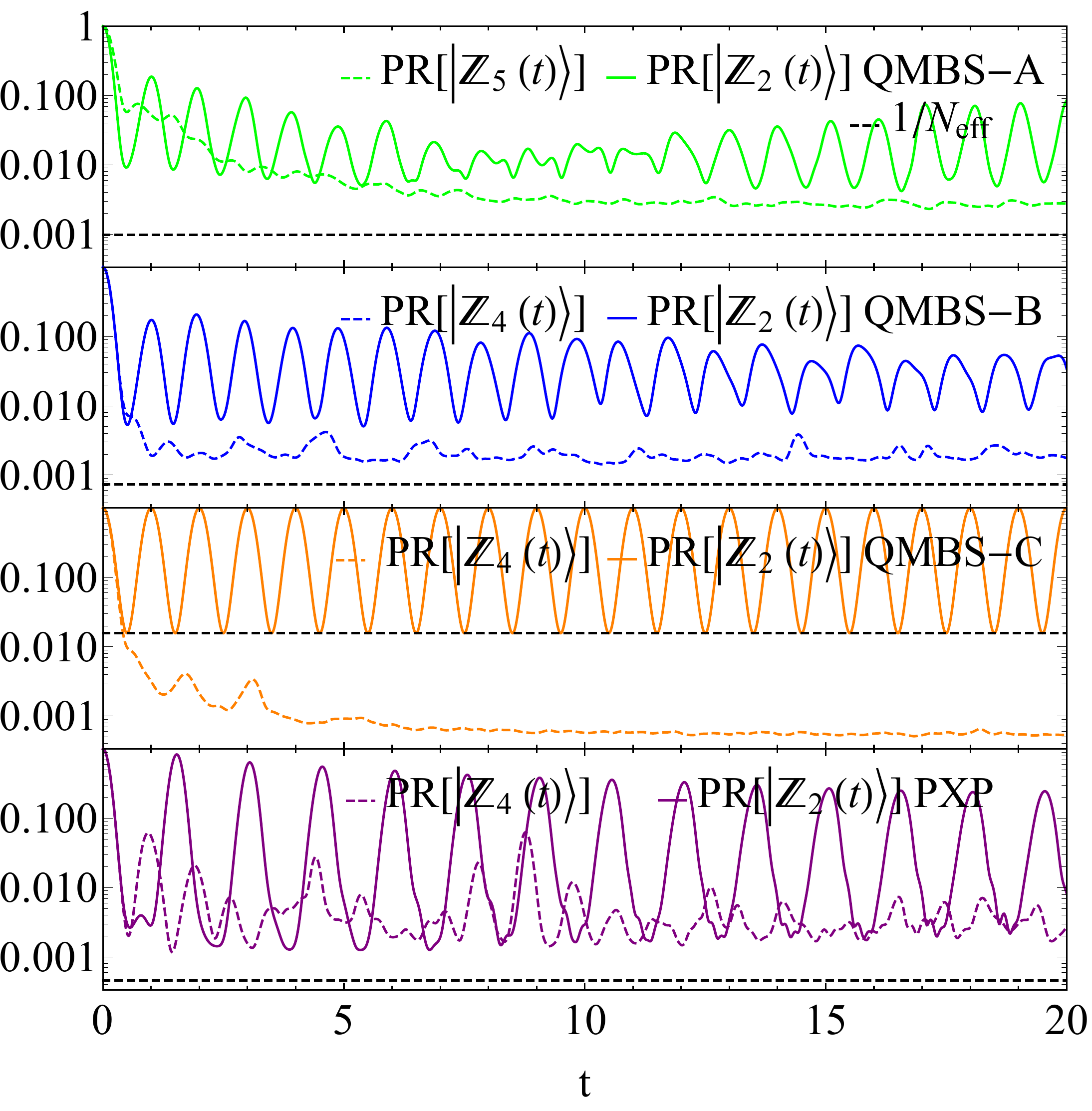}
    \caption[Revivals of the Néel states in the model studied in this work as seen from the PR.]{Revivals of the N\'{e}el state (solid line) and of a generic state (dashed line) showed on a log scale as seen from the PR of the time-evolved state for the various models studied. $L = 10,12,12,16$ and $N_{\text{eff}} = 1024,1366,64,2207$ for QMBS-A, QMBS-B, QMBS-C and PXP respectively.}\label{fig:revivals}
\end{figure}

As intuitively expected, the number of type I rules that are satisfied is correlated with the strength of the revivals. For instance, QMBS-B shows stronger, longer lasting and more coherent revivals compared to QMBS-A as can be seen in Fig.~\ref{fig:revivals} 
where the revivals are studied by considering the participation ratio (PR) of the time-evolved state $e^{-i(A + B)t}\ket{\mathbb{Z}_2}$ where $\ket{\mathbb{Z}_{k}}=|\underbrace{0, 1, 1,  \ldots 1}_{k},\underbrace{0, 1, 1,  \ldots 1}_{k}\ldots  \rangle$. The PR is evaluated in the basis of computational basis states  $\ket{q}$ and is defined as $\text{PR} \left[\ket{\psi} \right] = \sum_{q=1}^{2^L} \abs{\braket{q|\psi}}^4$ where $\ket{\psi}$ is assumed to be normalized. A PR close to $1$ indicates that the system is largely in one computational basis state while a PR $\sim 1/N_{\text{eff}}$, where the effective dimension $N_{\text{eff}}$ is defined here as the number of computational basis states connected to the Néel state by a matrix elements of some given power of $H$ (for the exact value of $N_{\text{eff}}$ in all the models studied, see App.~\ref{app:RstatSymmetrySector}), implies relaxation 
For comparison, the revival of a computational basis state that is not a Néel state is showed in Fig.~\ref{fig:revivals}, in which case it can be seen that the state quickly thermalizes. The exact scar model QMBS-C supports a spectrum generating algebra like many other exact QMBS models, and can also be viewed as an exact embedding which is discussed in App.~\ref{app:ExactEmbedding}. 

\begin{figure}
    \centering
    \includegraphics[width=0.46\textwidth]{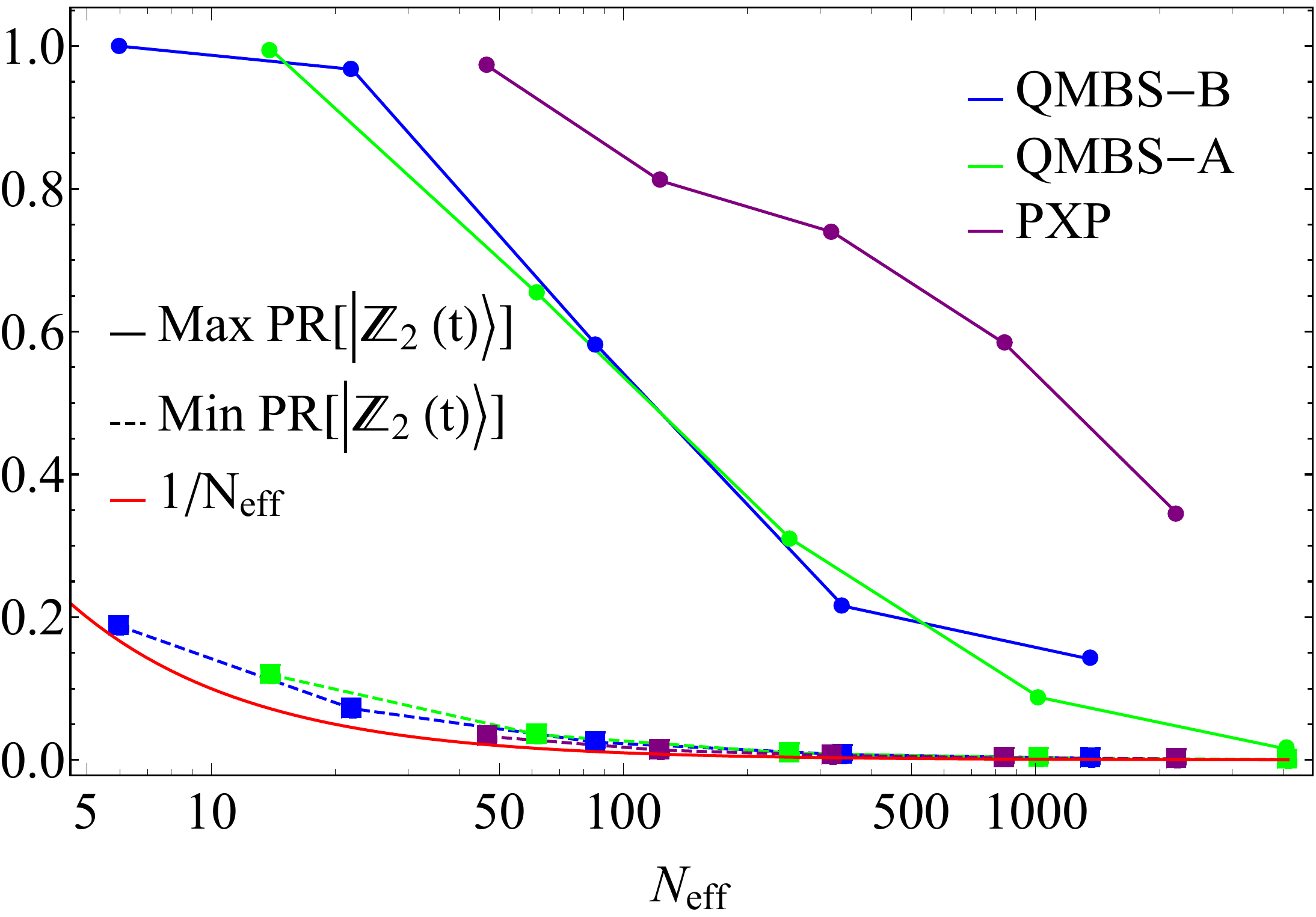}
    \caption[The maximum and minimum PR of the time evolved Néel states in the time range $t \in  (10, 300)$ versus the effective Hilbert space dimension $N_{\text{eff}}$]{The maximum and minimum PR of the time evolved Néel states in the time range $t \in  (10, 300)$ versus the effective Hilbert space dimension $N_{\text{eff}}$. The minimum closely follows the inverse effective Hilbert space dimension (red line) for all models. Satisfaction of more rules of type II/I appears to produce revivals that scale better with system size. }\label{fig:FiniteizeScaling}
\end{figure}

\begin{figure}
\centering
\includegraphics[width=0.46\textwidth]{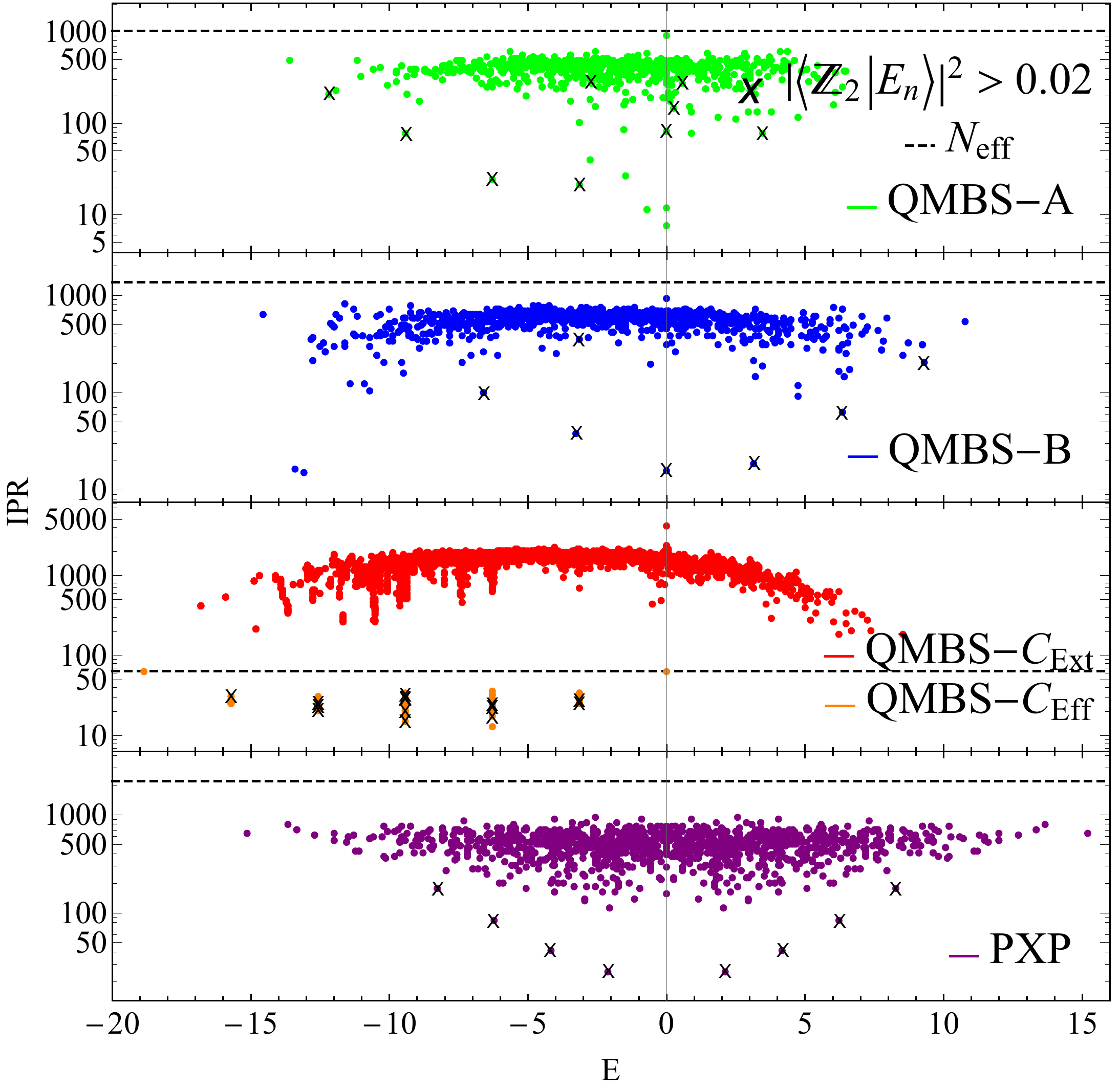}
\caption[Scatter plot of IPR vs. eigenstate energy]{Scatter plot of IPR vs. eigenstate energy in the 4 models studied in the main text. The states marked by an $x$ are eigenstates that have an overlap amplitude with $\ket{\mathbb{Z}_2}$ larger than $0.02$. Such states appear to be approximately equally separated in energy for all the models QMBS-A/B/C which is a hallmark of quantum scarring. Similar behavior is observed in the PXP model as well noted in Ref~\cite{turner2018quantum}. The scar signatures appear to be more pronounced provided a larger number of rules are satisfied. The Hamiltonians used to compute the eigenstates are restricted to the computational basis states appearing in the Kyrlov subspace associated with the Néel states, except for QMBS-C for which the full Hamiltonian is used to illustrate the embedding. Red dots in the QMBS-C panel show the IPR vs. energy of the eigenstates outside the Kyrlov subspace, whilst the orange dots show the IPR vs. energy of the eigenstates inside the Kyrlov subspace. $N_{\text{eff}} = 2207, 1024,1366,64$, $L = 16,12,10,12$ for PXP, QMBS-A, QMBS-B, QMBS-C respectively. See Sec.~\ref{sec:EffectiveDimension} for a precise definition of $N_{\text{eff}}$ }
\label{fig:IPRE}
\end{figure}

The presence of quantum scars in the models QMBS-A/B/C can also be seen from distribution plots of the inverse participation ratio $\text{IPR}[\ket{\psi}] = 1/\text{PR}[\ket{\psi}]$ of the eigenstates of the Hamiltonian $H = A + B$ for each model against their eigenenergies, as shown in Fig.~\ref{fig:IPRE} which reveals the presence of low IPR states that are exactly equidistant in energy for the QMBS-C model and approximately equidistant in energy for the PXP, QMBS-B and QMBS-A model. Furthermore, we also identify states which have a large overlap with the N\'{e}el states; these appear to coincide with the low IPR eigenstates (indicated by black $x$ in the figure). This strongly indicates a correlation between the number of rules of type I/II satisfied in the models and the presence of low $\text{IPR}$ states (scar eigenstates) in the spectrum. Finally, a finite-size scaling of the revivals in the PR of the time-evolved N\'eel states is shown in Fig.~\ref{fig:FiniteizeScaling}. The minima appears to coincide well with the inverse effective Hilbert space dimension $\sim 1/N_{\text{eff}}$, indicating near complete relaxation at intermediate times. The maxima corresponding to revivals, on the other hand, decreases with increasing system size but only as $-\text{log} (N)$ suggesting that the phenomena should be robust in the large $L$ limit.

\subsection{R-statistic and effective Hilbert space dimension}\label{sec:EffectiveDimension}
\begin{figure}
\centering
\includegraphics[width=0.46\textwidth]{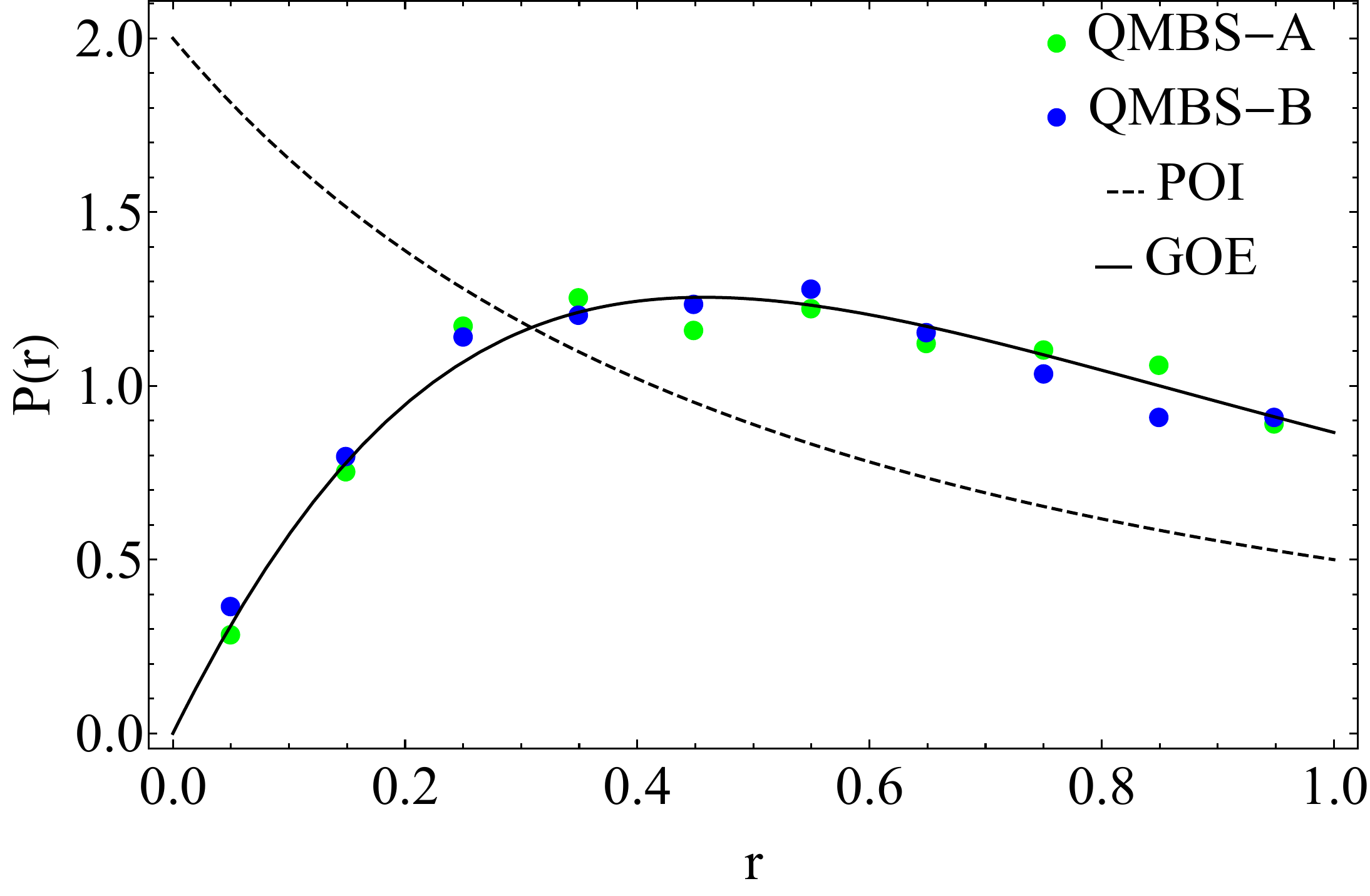}
\caption[Non-integrability of model QMBS-A and QMBS-B as seen from the suppression of P(r) at small r values.]{Non-integrability of model QMBS-A and QMBS-B can be seen in the suppression of P(r) at small r values. Eigenvalues for QMBS-A and B are computed in the basis of computational basis states that appear in the Kyrlov subspace associated with the Néel states. Furthermore, the Hamiltonian is restricted to the common +1 eigenspace of $S^2$ and $U_{SM}$ for $L = 16, 18$ which yields $4115$ and $4863$ eigenvalues for QMBS-A, QMBS-B respectively. }\label{fig:RStat}
\end{figure}

The level repulsion statistic, obtained as the ratio of the minimum to the maximum energy differences between successive eigenstates, $r_n = \text{min}(\Delta E_{n+1}/\Delta E_{n}, \Delta E_n/\Delta E_{n+1})$ where $\Delta E_n = E_{n} - E_{n-1}$, $E_{n} \leq E_{n+1}$, can be used as a metric to determine if a given model is integrable or not, which is key to showing that the approximate scars presented here are not due to integrability. By computing all the $r_n$ values for a given set of eigenvalues (extracted from a given symmetry sector of $H$) and constructing the associated probability density $P(r)$, one expects $P(r)$ to be Poissonian if the model is integrable, and charateristic of GOE/GUE ensembles if the model is non-integrable~\cite{RStatDistribution}. The most prominent feature of $P(r)$ for non-integrable models is suppression of $P(r)$ at $r$ values near $0$ which indicates level repulsion, a characteristic feature of non-integrable models. One can see in Fig. ~\ref{fig:RStat} that the models QMBS-A and QMBS-B show strong level repulsion and appear to closely follow GOE predictions indicating that they are non-integrable which rules out integrability as the reason for the presence of quantum scars in the models. For a detailed discussion of the symmetry sector (containing the scar states) studied, see App.~\ref{app:RstatSymmetrySector}.
\section{BCH expansion and revivals}\label{sec:BCHExpansionAndRevivals} An important natural question in our construction is how accurately the truncated Hamiltonian $H = A+B$ captures the dynamics we expect from the associated automaton unitary $U_F = e^{-iA} e^{-iB}$. In particular, of key interest to us is ensuring that the truncated Hamiltonian captures the dynamics of the automaton in the \emph{scar subspace}. It is clear that this is the case if all terms in the BCH expansion, $C_n$, vanish on the scar subspace. Since we do not know what this subspace is exactly in our construction, instead we examine the action of $C_n$ on the subspace of orbit states that our construction is designed to embed on to the scar subspace---recall the projector onto this subspace is denoted by $P_0$. 

In what follows, we will examine the typical matrix element of $C_n$ as a function of the order of the BCH exansion $n$, connecting i) orbit states to other orbit states, ii) orbit states to generic states, and iii) generic states to other generic states. We will examine these terms by numerically computing $\norm{P_0 C_n P_0}/l$, $\norm{(1-P_0)C_n P_0}/(l N_{\text{eff}})^{1/2}$ and $\norm{(1-P_0)C_n (1-P_0)}/N_{\text{eff}}$, respectively. Here, $\norm{X}$ denotes the Frobenius norm of the matrix $X$, and we divide this norm by $N_{\text{eff}} - l \approx N_{\text{eff}}$ (the Hilbert space dimension of generic states), or $l$ (the Hilbert space dimension of orbit states) or a composite of the two to obtain the value of the typical matrix element. 

We note apriori that ultimately, we would like the truncated Hamiltonian $H = A+ B$ to mimic the dynamics of the Floquet automaton on a putative scar subspace on which the selected orbit states have significant overlap. Although this is true when all matrix elements of BCH terms $C_n$ connecting scar states to generic states vanish, it is not obvious that examining the magnitude of terms in the BCH expansion is always the correct way of probing this aspect of the dynamics. For one, it may be the case that the BCH expansion may be reorganized in a way that appropriate linear combinations of $C_n$ have small matrix elements connecting orbit states to generic states even though individually the $C_n$ themselves have fairly large matrix elements. Second, here we attempt to examine the matrix elements between orbit states and generic computational basis states---even if these matrix elements are significant, it does not preclude the possibility that matrix elements of $C_n$ between scar \emph{eigenstates} of $H = A+ B$ and other generic states have small amplitude. The latter depends on how well the scar eigenstates actually embed the intended orbit states. We will see that in the PXP model, where rules of type II are satisfied, the BCH expansion does indeed show suppression of matrix elements between orbit states and generic states, order by order. With this clarification, we can now discuss our numerical findings.  

\subsection{Amplitude of BCH terms and possible prethermal behavior in the PXP model}

In Fig.~\ref{fig:BCHTermsAmplitude}, we plot the typical amplitude of the matrix elements of the $n^{\text{th}}$ order BCH term $C_n$ connecting various states in the Hilbert space. The following observations can be made---i) BCH terms connecting orbit states are heavily suppressed in the perfect scar model QMBS-C, and the PXP model, while they are suppressed only at certain specific orders in QMBS-A and QMBS-B, ii) matrix elements connecting orbit states to generic states decrease with $n$ at first for the PXP model, before eventually increasing again, iii) in the PXP model, even matrix elements connecting generic states to other generic states surprisingly show this phenomenology, iv) for QMBS-A/B, matrix elements connecting the scar subspace to generic states are smaller but of a similar magnitude to matrix elements between generic states, and v) in QMBS-C, the matrix elements connecting orbit states to generic states vanish exactly; this is to be expected as this is an exact scar model. 

Even though QMBS-A/B states show strong revivals only in the chosen orbit states, an order by order examination of terms in the BCH expansion does not reflect this fact---indeed, the matrix elements between orbit states and generic states is of the same order as those connecting generic states. As alluded to above, it may be possible to reorganize the BCH expansion in terms of linear combinations of various $C_n$, such that we do see suppression of matrix elements (between orbit states and generic states). We have not attempted this, but note that a natural reason for the failure of BCH expansion to capture this phenomena may be because these models were designed to strongly obey rules of type I---breaking this rules implies that for some set of powers, the local unitaries corresponding to the automata do not commute; see Eq.~(\ref{eq:typeIrules}). Since the local Hamiltonian is constructed as a linear combination of all powers of these local unitaries [Eq.~(\ref{eq:logUexpansion})], the BCH terms will be non-zero at all orders as soon as any of the type I rules (defined by the set of powers of the local unitaries) are broken. Note that in QMBS-C, all rules of type I are satisfied, it is an exact scar model, and it is thus not surprising that BCH terms at all orders have no matrix element connecting the scar states and generic states. 

The PXP model is different in that rules of type II can be enumerated naturally for this model, given the rather simple form of the local Hamiltonian term, and most of these rules are satisfied. As a result, we expect the BCH expansion to be more useful in this case. Specifically, in the PXP model, the norm of BCH terms first \emph{decreases} with $n$ before eventually increasing. This is characteristic of the FM expansion for systems driven at high frequencies and which concomitantly possess a prethermal window over which an effective Floquet Hamiltonian can be obtained by truncating the FM expansion. We explore this in more detail next.

\subsection{Prethermal behavior in the PXP model}

An interesting phenomenon that can occur whenever a quantum system is driven is Floquet prethermalization, which describes a prethermal time window inside which the driven quantum system reaches a prethermal quasi-steady state before slowly drifting towards true equilibrium. In particular, the length of that prethermal time window goes as $e^{1/\tau}$ where $\tau$ is the driving period. Such a prethermal window is normally accompanied by the norm of BCH expansion terms $\norm{C_n}$ first decreasing with $n$, up to some order $n_0$, before increasing with $n$. The duration of the prethermal window is then $\mathcal{O}(e^{1/n_0})$. Such a pattern is naturally obtained in the case of high frequency driving, for instance when $U_F = e^{-iA\tau}e^{-iB\tau}$ for small $\tau$, such that the lowest BCH terms largely decrease in $n$ as $n\tau^n$. In a many-body setting, eventually, the number of terms in the commutator in $C_n$ blows up as $n!$, which ultimately supresses the decays from $\tau^n$ at $n_0 \approx O(1/\tau)$. As a corollary, one can truncate the BCH expansion to order $n_0$ and expect the truncated Hamiltonian to mimic the Floquet unitary dyamics up to times $\approx e^{n_0}$. In this case, $\tau = 1$, and one cannot expect a prethermal regime on account of the frequency of the drive. However, by enforcing the commutator of $A, B$ to vanish on a subspace, one may expect a similar decrease of the norm of BCH terms before an eventual increase. 

\begin{figure}
    \centering
    \includegraphics[width=0.46\textwidth]{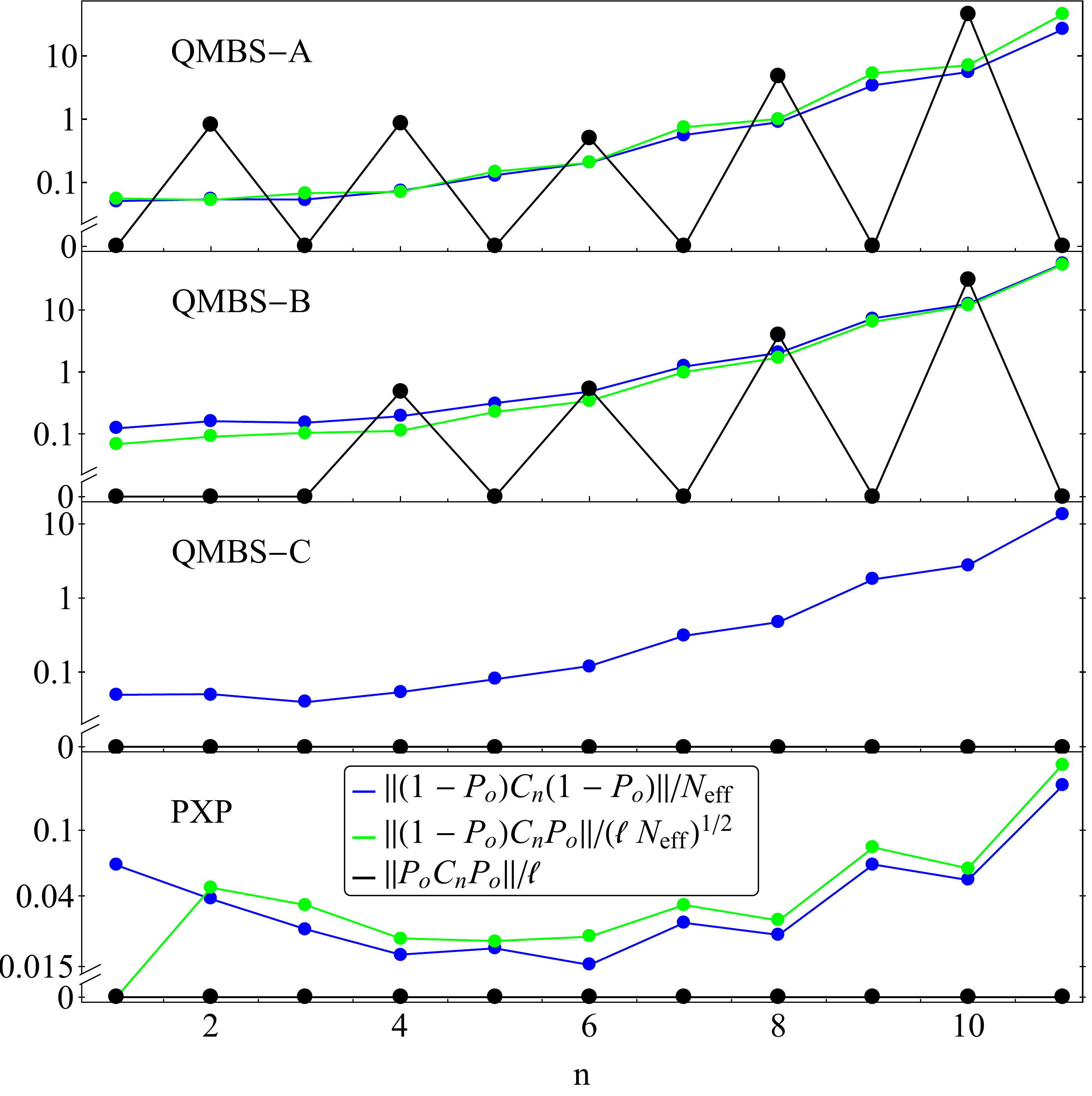}
    \caption[Leakage from orbit states, characterized by $\norm{(I-P_0)C_nP_0}$ and norm of the neglected terms $C_n$ projected to the subspace of generic-states and orbit states.]{ Leakage from orbit states, characterized by $\norm{(I-P_0)C_nP_0}$ (green) and norm of the neglected terms $C_n$ projected to the subspace of generic-states (blue) and orbit states (black). The amplitude of the BCH terms is normalized by the square-root of the number of matrix elements in the considered subspace, where $l$ is the length of the orbit to preserve and $N_\text{eff}$ is the number of computational basis states connected to the Néel state by a matrix elements of some given power of $H$. The $C_n$ are computed from the Hamiltonian terms $A$ and $B$ in the basis of computational basis states that appear in the Kyrlov subspace except for QMBS-C for which the calculation was performed on the full Hilbert space for illustrating the embedding.  $N_{\text{eff}} = 2207,4096, 1366$, $64$, $L = 16, 12, 12, 12$ for the PXP, QMBS-A, QMBS-B and QMBS-C model respectively. For the blue curve in QMBS-C, $N_{\text{eff}} = 4096 - 64$ is used.  }\label{fig:BCHTermsAmplitude}
\end{figure}

Indeed, as seen in Fig.~\ref{fig:BCHTermsAmplitude}, we do see that the amplitude of matrix elements connecting scar states to generic states decreases with the order of expansion $n$ before again increasing. Thus, there is an effective, emergent, time period $T_{\text{eff}} < 1$ which we may attribute to the fact that BCH terms $C_n$, which are composed of nested commutators of $A$ and $B$, are suppressed on the orbit subspace. Perhaps what is surprising is that the same behavior is in fact even seen for matrix elements between generic states in the computational subspace. 

The latter suggests that the prethermal dynamics may be applicable to not just the scar subspace, but to the full Hilbert space of the PXP model. To verify this, we examine the local autocorrelator,
%
%
%
%
 $|\braket{Z_i (t)} - \braket{Z_i}_{m.c}|^2$, where $\braket{Z_i}_{m.c}$ indicates the microcanonical average over a fixed energy window $\Delta E = 0.4$ centered around the average energy $E = \bra{\psi}H\ket{\psi}$ and $Z_i$ is the Pauli $\sigma_z$ operator acting on a particular spin $i$ of the system, which we choose arbitrarily. 
 
 Although many-body revivals of generic states decay rapidly, particularly in the PXP model, autocorrelations of local $Z_i$ continue to have long time revivals in any state. One may attribute this to the presence of a prethermal window---the dynamics of spins due to the underlying Floquet automaton show revivals, and within the prethermal window, this behavior is mimicked by the truncated, strictly local, Floquet Hamiltonian which in this case is the PXP model. 

\begin{figure}
    \centering
    \includegraphics[width=0.46\textwidth]{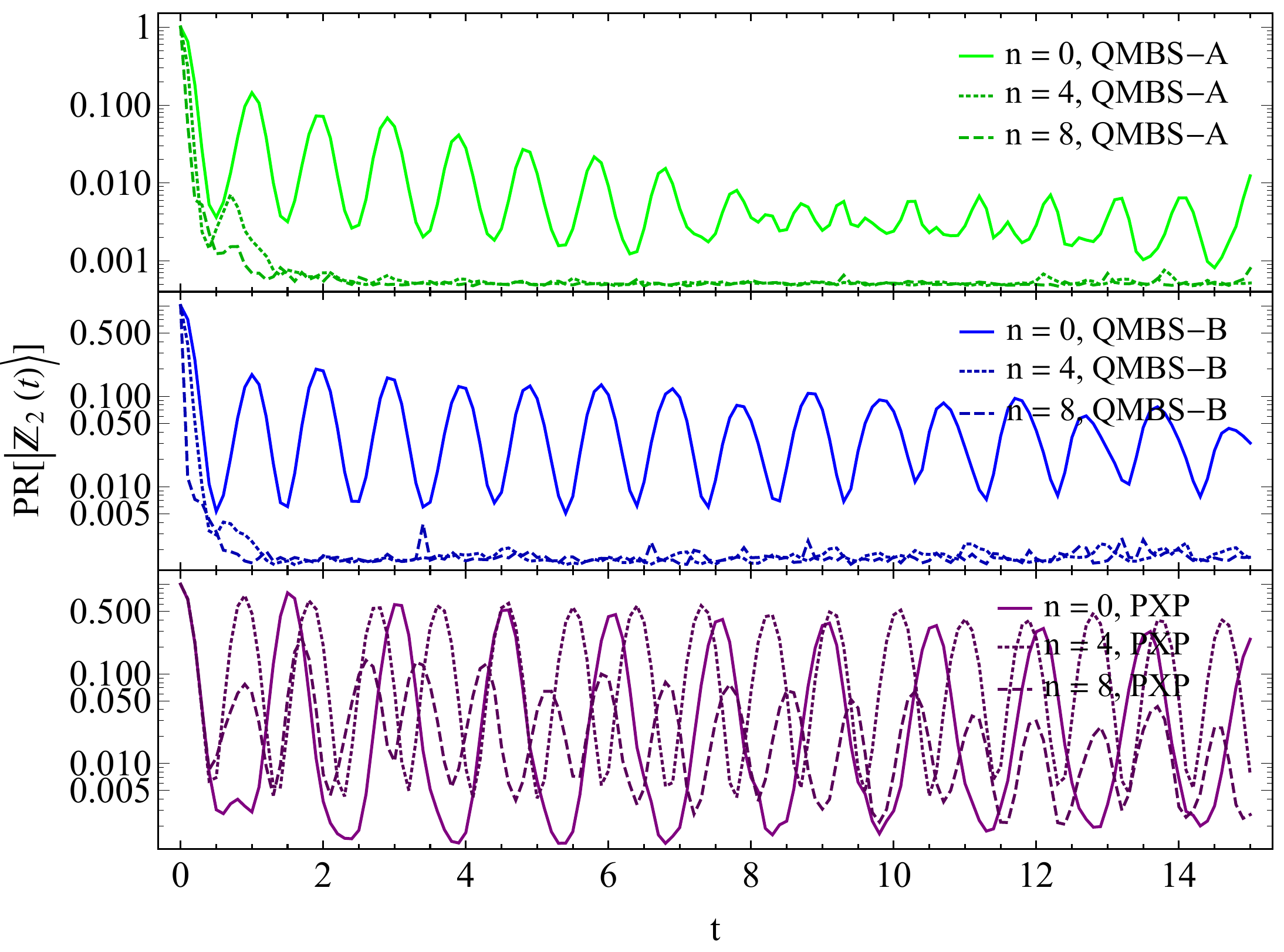}
    \caption[Revivals seen in models with additional BCH terms $C_n$ added to $A + B$.]{Revivals seen in models with additional BCH terms $C_n$ added to $A + B$. Revivals improve when adding up to fifth order of BCH terms in the PXP model but worsen upon adding further higher orders. In contrast, revivals only decrease when adding further terms in QMBS-A/B. $L = 12,12,16$ for QMBS-A, QMBS-B and PXP respectively.}
    \label{fig:IPRBCH} 
\end{figure}

To give further credence to this picture, we study the effect of adding the first few decreasing BCH terms to $H_{\text{eff}}$ $= A + B$ $= H_{\text{PXP}}$. We find that adding these terms in fact improves many-body revivals (both the revival strength and the regularity). Thus, one can think of the \emph{absence} of such terms in the truncated Hamiltonian $A+B$ as a perturbation away from the quasi-local Floquet Hamiltonian which captures the dynamics of the ideal Floquet automaton most faithfully; these terms lead to decay of revivals; see Fig.~\ref{fig:IPRBCH}.
\begin{figure}
    \centering
    \includegraphics[width=0.46\textwidth]{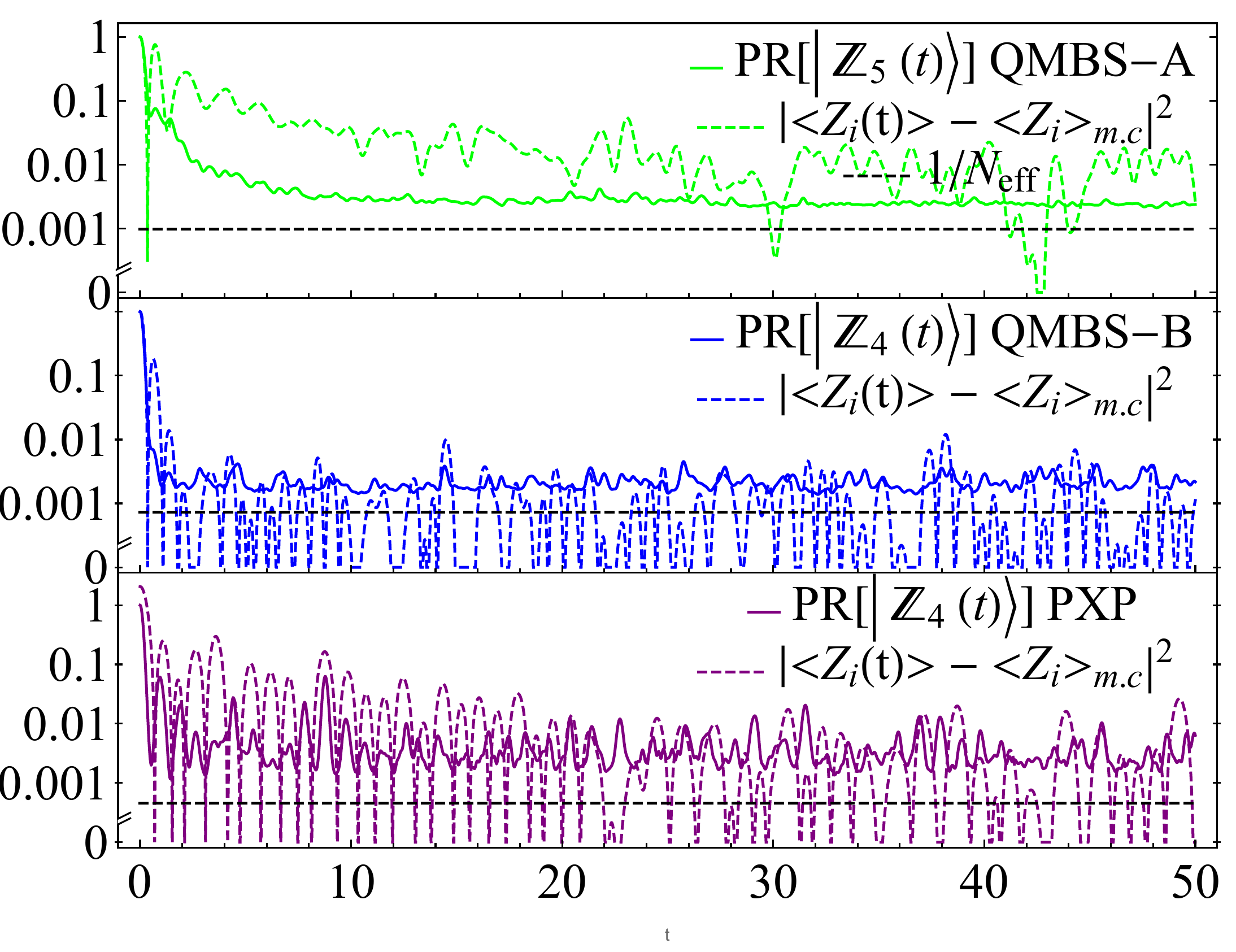}
    \caption[PR of generic states evolved in time and time evolution of $|\braket{Z_i(t)} - \braket{Z_i}_{\text{m.c}}|^2$]{PR of generic states evolved in time and time evolution of $|\braket{Z_i(t)} - \braket{Z_i}_{\text{m.c}}|^2$ where $\braket{Z_i}_{\text{m.c}}$ is the micro-canonical average computed with an energy window $\Delta E$ of 0.4 centred around the average energy of the considered generic state. QMBS-A and PXP show signs of prethermalization which manisfest themselves as a slow decay of $\braket{Z_i(t)}$ towards the micro-canonical average. }\label{fig:SigmazDecay}
\end{figure}
\\\\

Continuing with the analogy with Floquet systems driven at a high frequency and which exhibit a prethermalization window, we note the absence of terms $C_1, C_2, ..., C_{n_0 = 6}$ in our truncated Hamitonian $H_{\text{eff}} = A+ B$ can lead to decay of many-body revivals. We estimate this revival time by computing a Fermi's Golden Rule rate of decay of a scar eigenstate of $H_{\text{eff}} = A+B$ into non-scarred states. This rate is given by the typical matrix element $\Gamma$ in $C_2$ (which provides the largest coupling in the case of the PXP model; see Fig.~\ref{fig:BCHTermsAmplitude}) coupling this state to other states in the Hilbert space, multiplied by the number of states within an energy window $\Gamma$ around this chosen scar state $\sim \Gamma/\delta$, were $\delta$ is the many-body level spacing. The term $\norm{(1-P_0)C_2P_0}^2/(N_{\text{eff}}l)$ yields the norm squared  of a typical matrix element of the operator $C_2$. To estimate the many-body level spacing, we note that the scar eigenstate does not couple to all states in the Hilbert space. Some of the $C_i$ terms break full transnational symmetry and parity, but $S^2$, translation by two qubits, remains a conserved operator for all $C_i$. Thus, we can estimate the density of states $1/\delta$ within a given symmetry sector by $2N_{\text{eff}}/(L\Delta_{E_{\text{PXP}}})$ where $\Delta_{E_{\text{PXP}}} \approx 30$ is the bandwidth of the $\text{PXP}$ model for $L = 16$. The approximate decay rate is then given by $1/\tau \approx 2\pi(2N_{\text{eff}}/(L\Delta_{E_{\text{PXP}}}))\norm{(1-P_0)C_2P_0}^2/(N_{\text{eff}}l) \approx 0.1 $. This agrees approximately with an extrapolation of the numerical stimulated peak of many-body revivals to large times. 


(Note that for the PXP model, the terms within the N\'eel subspace are also small for reasons of locality, which prevents $C_n$ from leading to transitions between the N\'eel states) and symmetry (which prevents an energy offset between the N\'eel states due to particle-hole symmetry and translational symmetry in PXP). The latter likely aids stronger revivals and could be useful ingredients~\cite{Khemani2019,Choi2019} in searching for other approximate QMBS models using the methods outlined here.\\\\

\subsection{BCH terms in the PXP model}

We now examine these BCH terms for the PXP model in more detail. In particular, the first few orders are given by 
\begin{equation}
\begin{aligned}
C_0 + C_1 + C_2 = (-\frac{\pi}{2} +\frac{\pi^3}{96})\sum_jP_{j}X_{j+1}P_{j+2} \\
+ i\frac{\pi^2}{8}\sum_j (-1)^{j + 1}(P_{j} S_{j+1}^{+} S_{j+2}^{-} P_{j+3} - P_{j} S_{j+1}^{-} S_{j+2}^{+} P_{j+3}) -\\
\frac{\pi^3}{192}\sum_j ( P_{j}X_{j+1}P_{j+2}Z_{j+3}+Z_{j}P_{j+1}X_{j+2}P_{j+3}) + \\
\frac{\pi^3}{48}\sum_j(P_jS_{j+1}^+S_{j+2}^-S_{j+3}^+P_{j+4} + P_jS_{j+1}^-S_{j+2}^+S_{j+3}^-P_{j+4})
\end{aligned}
\end{equation}
, see App.~\ref{app:ExactPXPBCH} for a detailed derivation of this result.

Note that up to support over $4$ qubits, these corrections correspond to two terms, one which acts trivially on the orbit subspace, and the other, $( P_{j}X_{j+1}P_{j+2}Z_{j+3} +Z_{j}P_{j+1}X_{j+2}P_{j+3})$, was identified in both Refs.~\cite{Khemani2019,Choi2019} as a term that leads to better revivals and/or integrability of the model. This term was added to the PXP model with a variable amplitude which was optimized to improve integrability in Ref.~\cite{Khemani2019} and revivals in Ref.~\cite{Choi2019}. Here the magnitude of these terms is obtained without numerical optimization, and is given by that obtained from the BCH expansion. The ratio of the amplitude of this term to the PXP term is $\approx 0.129$, which is about $6$ times larger than that obtained in Ref.~\cite{Khemani2019} and $2$ times larger than that obtained in Ref.~\cite{Choi2019}. 

Finally, it is observed numerically that revivals in the PXP model improve upon adding BCH terms to an even order, while usually degrading upon adding terms to one additional order. This trend continues up to $n = 6$ after which revivals degrade with every successive order, see Fig.~\ref{fig:IPRBCH}. This can be attributed to renewed divergence of the BCH terms in the scar subspace beyond $n = 6$.

\subsection{PXP with and without phase}
\begin{figure}
\centering
\includegraphics[width=0.46\textwidth]{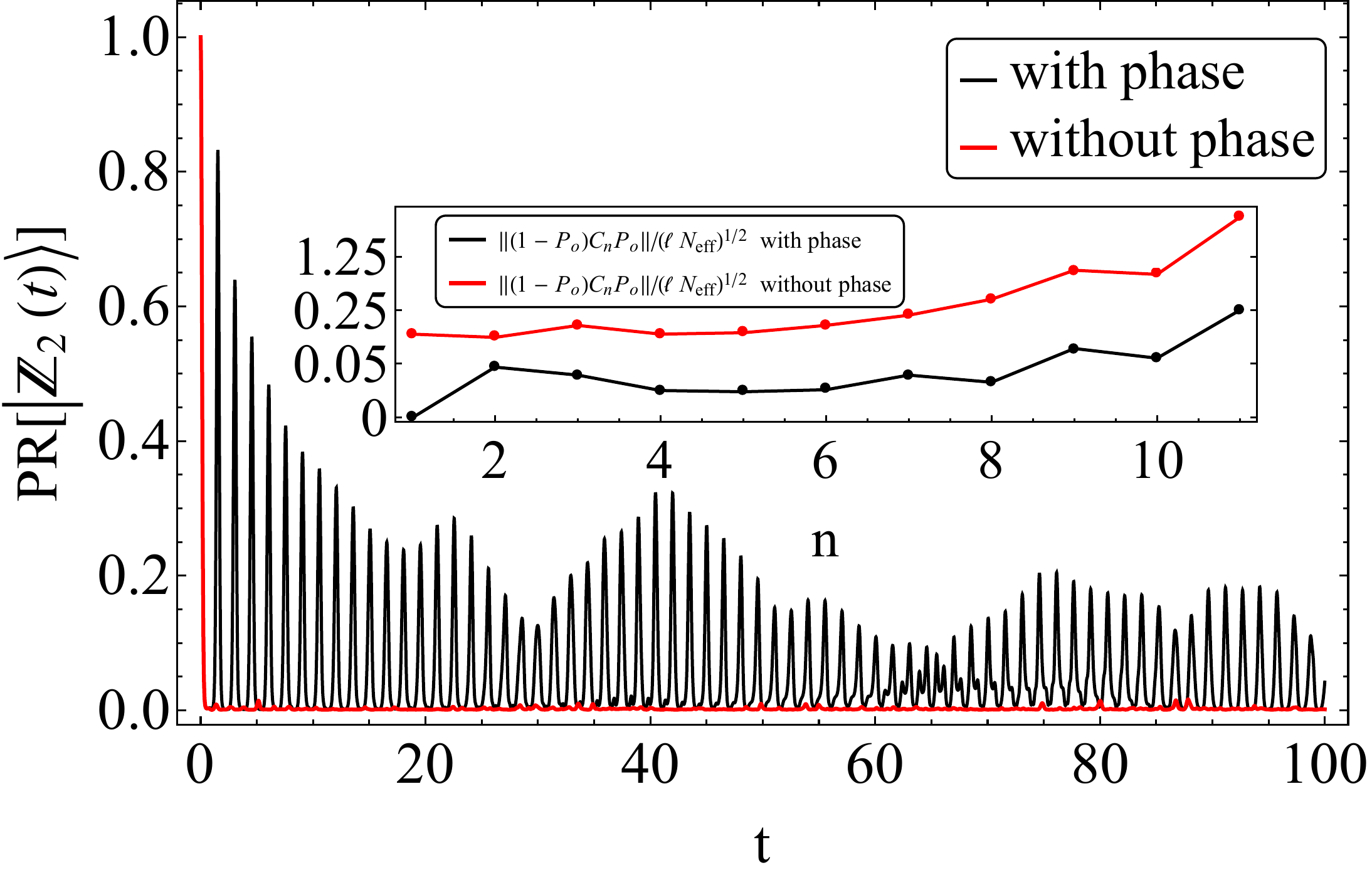}
\caption[Overlap amplitude squared of $\ket{\mathbb{Z}_2}$ with $e^{(A + B)t}\ket{\mathbb{Z}_2}$ for models where phase is trivial vs non-trivial.]{Overlap amplitude squared of $\ket{\mathbb{Z}_2}$  with $e^{-i(A + B)t}\ket{\mathbb{Z}_2}$ for models where phase is trivial vs non-trivial. Insets show how the effective leakage changes between models with trivial and non-trivial phase.}
\label{fig:phaseVsNoPhase}
\end{figure}

The phase that states accrue as they evolve under the Floquet automata can play a very important role. Fig.~\ref{fig:phaseVsNoPhase} highlights the stark difference in revival strength from automata with unitaries enforcing the same permutation but one in which the phase is trivial, and the second in which it is non-trivial. The second one corresponds to the usual PXP Hamiltonian. One can see that the former model exhibits smaller revivals which further corroborates the intuition that the amplitude of $\norm{(1-P_o) C_n (P_o)}$ is correlated with the strength of the revivals. For instance, the first leakage term fully vanish in the PXP model, but it dosen't in the related model. (Note that a continuum of models between these two extremes was studied in \cite{Mukherjee2020Tuning},\cite{Mukherjee2020Vacuum}.) 

\section{Discussion and outlook}\label{sec:DiscussionOutlook}
In this work, we show a method for engineering quantum many-body scar Hamiltonians by establishing a connection between quantum cellular automata $U_F = e^{-iA}e^{-iB}$ and the Hamiltonian $H = A + B$ obtained by carefully taking the matrix logarithm of each layer of $U_F$. Generally, the dynamics generated by $U_F$ and $H$ are unrelated. In particular, one can view $H$ as a Hamiltonian obtained by truncating the BCH expansion $U_F = e^{-i (A+B) -i C_1 - i C_2 ...}$ to zeroth order. Although for a generic interacting system, these terms $C_n$ rapidly diverge, we devise two sets of rules, dubbed rules of type I and type II, that force these terms to vanish on a small subspace of states that are part of a cycle of $U_F$ of finite length $l$. 

We then construct models QMBS-A/B/C that successively satisfy more of the local rules of type I, which enforce that certain local commutators of the unitaries in $U_F$ vanish on a cycle composed of the two N\'{e}el states. The PXP model is more naturally interpreted as satisfying a large number of rules of type II; these rules enforce that local commutators of the Hamiltonian vanish on a cycle of length $3$ composed of the two N\'{e}el states and a vacuum state. The models QMBS-A/B/C satisfy successively more rules and exhibit concomitantly stronger revivals, with QMBS-C being an exact scar model. 

We also examined, order by order, the typical matrix element in $C_n$ connecting scar states to generic states. The amplitude of these terms is ideally heavily suppressed as it causes leakage from the scar subspace to generic states. We find that these terms decrease with increasing $n$ in the PXP model, before again begininning to diverge at order $n_0 = 6$. This behavior is characteristic of prethermalization phenomena in Floquet systems driven at high frequencies. Although in this case the drive frequency is putatively $1$, the observed behavior of the BCH terms suggests an emergent timescale $T_{\text{eff}} < 1$ and a prethermal window $\tau_p \sim e^{1/T_{\text{eff}}}$. In fact, even matrix elements of BCH terms connecting different generic states in the PXP model appear to show the same behavior. We find evidence of prethermal behavior in the PXP model by looking at autocorrelators of local spin $Z_i$. This operator shows revivals even in generic states, and long after many-body revivals of this generic state have decayed. We can also recover a timescale for decay of many-body revivals by computing a Fermi's Golden rule rate of decay based on the amplitude of matrix elements connecting orbit states to generic states in the BCH terms neglected. 

The BCH expansion does not appear to be the correct way to understand leakage out of the scar subspace (and thus, many-body revivals) in the case of QMBS-A/B. Here, order by order, matrix elements connecting scar states to generic states can be of the same order as those connecting different generic states. This seems to contradict the fact that the orbit states are special and distinct from generic states because only these states show many-body revivals. A natural explanation is that the BCH expansion may be reorganized in a way such that linear combinations of various $C_n$ may have a small matrix element connecting orbit states to generic states even though individually the $C_n$ have sizeable matrix elements. This requires further exploration. 

We note that we may interpret the results of our work without directly appealing to Floquet automata. The rules devised effectively ensure that a putative scar Hamiltonian $H$ can be decomposed into a partition $A + B$, where scar states are common eigenstates of $A$ and $B$. If $A$ and $B$ are composed of physically disjoint terms, they naturally possess eigenstates of low entanglement. If $e^{i n A} = e^{i n B} =\mathbb{1}$, for some integer $n$, the eigenvalues of $A$ and $B$ are equidistant. Ensuring that commutators of $A,B$ vanish on a certain (scar) subspace ensures that there exist a limited number (scaling at most polynomially in the system size $L$) of common eigenstates of $A, B$ that are equidistant in energy. If we can somehow embed further low-entanglement states in this subspace, as we do, then one obtains a scar subspace of low-entanglement eigenstates.

In many ways, this work is a first step in leveraging the properties of non-thermal quantum cellular automata to construct quantum many-body scars. Some questions emerge naturally from this work. For instance, the choice of partitioning of the Hamiltonian into two parts $A$ and $B$ where $A$ concerns the `even' gates, and $B$ the `odd' gates is rather arbitrary. Nothing prevents one from choosing a different decomposition of $H$ which would ultimately lead to a distinct automaton being associated with $H$. Provided this new automaton satisfies all or a large number of local rules for some specific states $\ket{\psi}$, it might be possible to identify additional quantum scar towers in the same model. We note here that in related work, we show that the mid-spectrum scar states in the AKLT model for instance can be obtained by considering various partitions of this model. 
Another interesting avenue for future work would be to study local unitary gates $U_{0,j}$ that are not simple permutation gates with phase. The construction presented here in principle applies to any unitary gate $U_{0,j}$ that satisfies the property $U_{0,j}^n = \mathbb{1}$ for some integer $n$ regardless of the internal structure of $U_{0,j}$. Such an approach might lead  to quantum scars with more complex structures. An interesting avenue for doing so would be to consider Clifford gates as the local unitary gates $U_{0,j}$ which, despite generating entanglement are entirely described by an underlying classical automaton which acts as a permutation of the set of products of Pauli matrices rather than the computational basis states themselves. 
\\\\

\section{Acknowledgements} The authors acknowledge useful discussions with Sarang Gopalakrishnan, Thomas Iadecola, Sanjaya Modugalya and Shivaji Sondhi. KA would like to acknowledge the hospitality of the Aspen Center for Physics where some of the ideas in this work were developed. PGR acknowledges graduate funding support from NSERC and FRQNT. MJG acknowledges support from  the National Science Foundation (QLCI grant OMA-2120757). KA acknowledges funding support from NSERC, FRQNT, and the Tomlinson Scholar Award. 

\appendix

\section{Permutation and phase map representation}\label{app:PermPhaseDef}

In order to characterize the unitary matrices $U_{0,j}$ and their properties, it is convenient to introduce a compact way of representing them. Since the $U_{0,j}$ act as a permutation on the computational basis states on which they act as well as multiplying them by a phase, they can be represented using the cycle notation of a permutation as well as a phase map. One can associate the computational basis states with integers between $1$ and $16$ by converting their base 2 bit string representation to an integer, $+1$. Explicitly,
\begin{equation}
\begin{aligned}
    \ket{0000} \rightarrow \ket{1}\quad 
    \ket{0001} \rightarrow \ket{2} \text{ }\ldots\text{ }
    \ket{1111} \rightarrow \ket{16}.
\end{aligned}
\end{equation}
The state $\ket{0}$ is understood to be the $+1$ eigenstate of $2S_z$ and the $\ket{1}$ state is the $-1$ eigenstate of $2S_z$ where $S_z$ is the standard $z$ spin operator for a spin $1/2$ particle. The phase map is represented by an array of length 16 $(\text{ph}_1, \text{ph}_2,...,\text{ph}_{16})$ with the understanding that the $q^{\text{th}}$ component $\text{ph}_q$ of this array is the complex number by which the $q^{\text{th}}$ computational basis state is multiplied when acted upon by $U_{0,j}$, i.e $U_{0,j}\ket{q} = \text{ph}_q\ket{\sigma(q)}$, see Fig.~\ref{fig:U0Permutation}. The transitions between computational basis states are represented with the cycle notation of a permutation, e.g if the permutation matrix $U_{0,j}$ generates the transitions $(1\rightarrow 3 \rightarrow 8 \rightarrow 1)$, $(2\rightarrow 4 \rightarrow 2)$ and sends all other states to themselves (possibly with a phase), then one can compactly represent the above transitions by $((1,3,8),(2,4))$ where it is understood that consecutive integers $n_i, n_{i+1}$ in a cycle $(n_1,n_2,...,n_l)$ represent a transition from $n_i$ to $n_{i+1}$. The cycle is periodic in the sense that the last integer that appears in the cycle denoted above by $n_l$ is mapped to $n_1$. Any computational basis state that do not appear in a cycle is assumed to be mapped to itself.

\section{Total number of relevant rules}\label{app:TotNumberRules}
Some of the rules that appear in Eq.~\ref{eq:typeIrules} are trivially satisfied. Indeed, for the rules to be non-trivial, it must be the case that $s_2$ is non zero, and that at least one of $s_1$ or $s_3$ is non-zero. Most generally, this yields a total of $l(n-1)(n^2 - 1)(L/2)$ rules where $l$ is the length of the cycle to be preserved and $L$ is the system size. If the states $\ket{\sigma^n(q)}$ spawning the subspace to be protected are such that $S^2\ket{\sigma^n(q)} = \ket{\sigma^n(q)}$ where $S$ is the operator translating all sites by one to the right, then the number of relevant rules is reduced to $l(n-1)(n^2 - 1)$ and is independent of system size. In the remainder of this work, for a given system, the number of satisfied rules is presented as a fraction of the total number of relevant rules, i.e. it will be presented as $\text{(Number of satisfied rules)}$ $/(\text{Total number of relevant rules})$.

In the geometry where $U_F = \prod_{i}^{L/4}U_{0,4j-3}\prod_{j}^{L/4}U_{0,4j-1}$ and with $P_0$ composed of the two Néel states, one has a total number of relevant rules given by $l(n-1)(n^2 - 1)$ where $n = 6$, $ l = 2$, so a total of $350$ relevant rules. The PXP model is special since the local Hamiltonian has the property that $h_{0,j}^3 = \frac{\pi^2 }{4} h_{0,j}$, so it is worth considering rules of type II instead. The total number of relevant rules of type II for the PXP model is given by $(n-1)(n^2 - 1) + (n-1)(n^2 - 1)2$ with $n = 3$, so a total of 48 rules. The first term counts the rules associated with the fully polarized state $\ket{1111...}$, the second term counts all the rules associated with the state $\ket{1010...}$. Note that since the Néel states are such that $\ket{1010...} = S\ket{0101...}$, one directly obtains that satisfying all the rules for one of the two Néel states (taking into account that the Néel states are not translationally invariant) ensures that the rules are satisfied for the other Néel state as well, so no additional rules need to be taken into account.
\section{Decomposing Hamiltonian's in terms of powers of simple unitary matrices}\label{app:SolvingForC}
The coefficients $c_k$ that appear in Eq.~\ref{eq:logUexpansion} can be found by writing Eq.~\ref{eq:logUexpansion} with a set of orthonormal eigenvectors. Doing so, one obtains 
\begin{equation}
   \sum_{s=1}^{2^4}-\tilde{\beta_s}\ket{\beta_s}\bra{\beta_s} = \sum_{s = 1}^{2^4}(\sum_{k=1}^{n}e^{ik\beta_s}c_k)\ket{\beta_s}\bra{\beta_{s}}
\end{equation}
where $\ket{\beta_s}$ is an eigenstate of $U_F$ with eigenvalue $\beta_s$. This yields the matrix equation
\begin{equation}\label{conditionlogexpasion}
\begin{aligned}
M\vec{c} = -\vec{\beta}, \quad M_{s,k} = e^{ik\beta_s}, \\ s \in \{1,2,...,2^4\},\quad k \in \{1,2,...,n\}
\end{aligned}
\end{equation}
with $\vec{c} = (c_1,c_2,...,c_{n})$ and $\vec{\beta} = (\tilde{\beta}_1,\tilde{\beta}_2,...,\tilde{\beta}_{2^4})$ In this form, it is not obvious that Eq.~\ref{conditionlogexpasion} always admits a solution, but it turns out that a solution does indeed always exist.

To construct it, consider the following set of states 
\begin{equation}\label{eq:EigenstateExtension}
    \sum_{k = 0}^{n-1}e^{i\alpha_k}U_{0,j}^{k}\ket{q}
\end{equation}
for some set of real numbers $\alpha_k$. Note that since $n$ is such that $U_{0,j}^n = \mathbb{1}$, this sequence of states is a closed loop upon successive applications of $U_F$. It is easy to see that the choice $\alpha_{k} = -k\gamma$, where $\gamma$ is one of the $n$ roots of unity, yields an eigenstate of $U_F$ with eigenvalue $e^{i\gamma}$. Note that all the generated eigenstates produced by Eq.~\ref{eq:EigenstateExtension} for a given $\ket{q}$ have distinct eigenvalues and are thus orthogonal, but Eq.~\ref{eq:minimalEigenstates} suggests that one should only be finding $l$ eigenstates where $l$ is the length of the cycle. Furthermore, eigenstates built from different cycles are necessarily orthogonal to each other since they contain different computational basis states. For this to be possible, it must be the case that some of the eigenstates produced with \ref{eq:EigenstateExtension} are equal to the vector $\vec{0}$. One can deduce from the previous sections that the only non-zero eigenstates will be the ones for which $\gamma$ is given by $\gamma = \frac{\Phi + 2\pi m}{l}$, see Eq.~\ref{eq:ConditionEigenstates}. The redundant eigenstates can safely be added to the eigenstate decomposition of $U_{0,j}$ and $i\log{U_{0,j}}$ just like if they were non-zero vectors which is key to solving for the vector $\vec{c}$. Doing so yields
\begin{equation}
\begin{aligned}
   \sum_{s = 1}^{nN_{\text{Cycles}}}-\tilde{\gamma_s}\ket{\gamma_s}\bra{\gamma_s} =\\ \sum_{s = 1}^{nN_{\text{Cycles}}}(\sum_{k=1}^{n}e^{ik\gamma_s}c_k)\ket{\gamma_s}\bra{\gamma_s}
\end{aligned}
\end{equation}
where $\ket{\gamma_s}$ are eigenstates of $U_{0,j}$ with eigenvalue $e^{i\gamma_s}$ now also including the redundant eigenstates. $\tilde{\gamma_s}$ is equal to $-i$ times the principal logarithm of $e^{i\gamma_s}$ and $N_{\text{Cycles}}$ is the total number of cycles composing $U_{0,j}$. From this, one obtains the matrix equation
\begin{equation}
\begin{aligned}
\Gamma\vec{c} = -\vec{\gamma}\qquad\Gamma_{s,k} = e^{ik\gamma_s} \\ s \in \{1,2,...,nN_{\text{Cycles}}\}\qquad k \in \{1,2,...,n\}
\end{aligned}
\end{equation}
with $\vec{c} = (c_1,c_2,...,c_n)$, $\vec{\gamma} = (\tilde{\gamma}_1, \tilde{\gamma}_2,...,\tilde{\gamma}_{nN_{\text{Cycles}}})$.
Remarkably,  $\frac{\Gamma^{\dagger}}{nN_{\text{cycles}}}$ is an inverse of $\Gamma$ 
\begin{equation}\label{eq:solutionCoefficients}
\begin{aligned}
   \frac{(\Gamma^{\dagger}\Gamma)_{k,m}}{nN_{\text{cycles}}} =  \sum_{s=1}^{nN_{\text{cycles}}}\frac{\Gamma^*_{s,k}\Gamma_{s,m}}{nN_{\text{Cycles}}}\\ = \sum_{s=1}^{nN_{\text{cycles}}}\frac{e^{-i\gamma_s(k-m)}}{nN_{\text{Cycles}}} = \delta_{k,m} 
\end{aligned}
\end{equation}
To see why Eq.~\ref{eq:solutionCoefficients} is valid, note that the sum over $s$ runs over the augmented eigenvalues $\gamma_s$ associated with each cycle composing $U_{0,j} $. $k$ and $m$ both take values between $1$ and $n$, so their difference $k - m$ takes values in the range $[-n + 1, n-1]$. $\gamma_s$ is one of the $n$ roots of unity modulo $2\pi$. One can then decide to order the eigenvalues by choosing $\gamma_s = \frac{2\pi s}{n}$, $s \in \{1,2,...,nN_{\text{cycles}}\}$ where say the $n$ first eigenvalues are associated with the first cycle, the $n$ next with the second cycle, so on and so forth. Eq.~\ref{eq:solutionCoefficients} then reads
\begin{equation}
\begin{aligned}
    \frac{1}{nN_{\text{cycles}}}\sum_{s=1}^{nN_{\text{cycles}}}e^{-i\frac{2\pi s(k-m)}{n}}\\ = \frac{N_{\text{cycles}}}{nN_{\text{cycles}}}\sum_{s=1}^{n}e^{-i\frac{2\pi s(k-m)}{n}} \\= \frac{1}{n}(\frac{1-e^{2\pi(k-m)(n+1)/n}}{1 -e^{2\pi(k-m)/n}} - 1)
\end{aligned}
\end{equation}
which always yields 0 provided $(k - m)$ is not a multiple of $n$. As seen above, $(k - m)$ takes values in the range $[-n+1,n-1]$, so one obtains an indeterminate result only when $k = m$, in which case it can directly be seen that the result is 1. This implies that the coefficients $c_k$ are given by
\begin{equation}
   \vec{c} = \frac{\Gamma^{\dagger}}{nN_{Cycles}}\vec{\gamma} 
\end{equation}
which provides an explicit method for decomposing $i\log(U_{0,j})$ as a linear superposition of powers of $U_{0,j}$. Remarkably, the vector $\vec{c}$ only depends on the order of the unitary matrix $n$, so distinct unitary matrices with the same order $n$ assume the same decomposition.

\section{QMBS-C as an embedded spectrum generating algebra}\label{app:ExactEmbedding}
The full Hamiltonian corresponding to the model QMBS-C is given by 
\begin{equation}\label{eq:QMBSCFullH}
\begin{aligned}
    H = \sum_{j = 1}^{L/2}\left(\frac{\pi}{2}P_{2j}X_{2j}X_{2j+1}P_{2j} + \right.\\ \left.(1 - P_{2j})H_{\text{ext},2j-1}(1 - P_{2j}) -\frac{\pi}{2}I\right)
\end{aligned}
\end{equation}
with $P_{j+1} = (I - Z_{j+1}Z_{j+2})/2$. $H_{\text{ext},j}$ is given in Tab.~ \ref{tab:PXPTabSpin}, but the exact form of $H_{\text{ext},j}$ turns out to be irrelevant. Let's begin by showing that \ref{eq:QMBSCFullH} is an embedded model. Note first that the set of projectors $P_{2j}$ and the Hamiltonian all mutually commute, a state of the system can thus be an eigenstate of all $P_{2j}$ simultaneously. This fact allows one to directly connect the model \ref{eq:QMBSCFullH} to the embedding method presented in \cite{shiraishimori}. The arbitrary Hamiltonian terms $h_j$ correspond to $H_{\text{ext},j}$ and the Hamiltonian $H'$ is $\sum_{j}\frac{\pi}{2}P_{2j}X_{2j}X_{2j+1}P_{2j}$. For a state $\ket{\psi}$ to be a $+1$ eigenstate of the $P_{2j}$, it must be the case that qubits sitting on sites $2j, 2j+1$ have opposite spin. The subspace spawned by such states has dimension $2^{L/2}$ and includes for instance the two Néel states. The effective Hamiltonian acting on this subspace is given by  
\begin{equation}\label{eq:QMBSCHeff}
    H_{\text{eff}} = \sum_{j = 1}^{L/2}\left(\frac{\pi}{2}X_{2j}X_{2j+1} -\frac{\pi}{2}I\right)
\end{equation}
which is obtained by setting all $P_{2j}$ to $I$. The full Hamiltonian H hosts a spectrum generating algebra, see Ref. \cite{markaklttowers2020} for an introduction to the topic. Indeed, consider the operator 
\begin{equation}\label{eq:}
    Q^{\dagger} = \sum_{j=1}^{L/2}Z_{2j}(I - X_{2j}X_{2j + 1})
\end{equation}
and consider the linear subspace W spawned by the $2^{L/2}$ states that are in the common $+1$ eigenspace of the $P_{2j}$. This operator can be seen to be responsible for a spectrum generating algebra. Indeed one has that 
\begin{equation}
    ([H,Q^{\dagger}] - \epsilon Q^{\dagger})W = 0
\end{equation} which follows from
\begin{equation}
\begin{aligned}
\relax[H,Q^{\dagger}]W \\= \sum_{j=1}^{L/2}[\frac{\pi}{2}P_{2j}X_{2j}X_{2j+1}P_{2j},Z_{2j}(I - X_{2j}X_{2j + 1})]W \\= \pi\sum_{j=1}^{L/2} Z_{2j}(I - X_{2j}X_{2j + 1})W = \pi Q^{\dagger}W
\end{aligned}
\end{equation}
where the second equality comes from $(I - P_{2j})W = 0$, $P_{2j}W = W$ and $Q^{\dagger}W \subset W$. One can see from the above that $\epsilon = \pi$. This shows that QMBS-C hosts a spectrum generating algebra. QMBS-C is thus an example of a model where one observes an embedded spectrum generating algebra in an otherwise fully thermal Hamiltonian.

\section{Symmetry sectors of QMBS-A/B}\label{app:RstatSymmetrySector}
The relevant symmetries of the QMBS-B and QMBS-A model are invariance under $S^2$ and invariance under the unitary operator 
\begin{equation}
    U_{SM} = \left(\prod_{i=1}^LX_i\right)SM
\end{equation}
where $S$ is the operator that shifts all sites by one to the right and $M$ is the mirror operation about the center bond. Furthermore, both models posses the anti-unitary symmetry $RSM$ where $R$ is the complex conjugation operation, which implies time reversal symmetry (which explains why the GOE ensemble is the best fit for $P(r)$). In order to compute the R-statistic, one must restrict the Hamiltonian to a given symmetry sector, which is chosen in this case to be the common $+1$ eigenspace of $S^2$ and $U_{SM}$, which for instance contains the state $\frac{1}{\sqrt{2}}(\ket{1010...}+\ket{0101...})$. Furthermore, one must also restrict the Hamiltonian to the set of computational basis states that appear in the Kyrlov subspace associated with the Néel states. For QMBS-B, one can see that the local Hamiltonian $h_{0,j}$ can only lower or increase the total spin $Z_{\text{tot}} = \sum_iZ_i$ by multiples of three, see Tab.~ \ref{tab:PXPTabSpin}. Thus, the total number of accessible computational basis states starting from the Néel state is given by the set of all computational basis states that have a total spin $Z_{\text{tot}}$ which is separated from $Z_{\text{tot}} = 0$ by some multiple of 3 (the Néel states are such that $Z_\text{tot} = 0$). In QMBS-A, no such restrictions exists. In QMBS-C, the total number of accessible states from the Néel states is given by $2^{L/2}$, see App.~\ref{app:ExactEmbedding} for a more precise definition. Finally, the PXP model is restricted to the well known Fibonacci subspace \cite{turner2018quantum}. The effective dimension $N_{\text{eff}}$ is defined here as the number of computational basis states connected to the Néel state by a matrix elements of some given power of $H$. It is given here for all the models studied in this work.
\begin{equation}
    \begin{aligned}
    N_{\text{eff},\text{PXP}} = F_{L+1} + F_{L-1}\\
    N_{\text{eff},\text{QMBS-A}} = 2^L\\
    N_{\text{eff},\text{QMBS-B}} = \sum_{k = -\lfloor{L/3}\rfloor}^{\lfloor{L/3}\rfloor}\frac{L!}{(\frac{L}{2} + 3k)!(\frac{L}{2} - 3k)!}\\
    N_{\text{eff},\text{QMBS-C}} = 2^{L/2}
    \end{aligned}
\end{equation}
where $F_n$ is the $n^{\text{th}}$ Fibonacci number and $L$ is the system size. Note that the above is only well defined for even system sizes, the Néel states do not exist otherwise.

\section{Exact PXP BCH terms}\label{app:ExactPXPBCH}
The first order BCH term $C_1$ is given by
\begin{equation}
     -\frac{i}{2}[A,B] = \sum_{j\in \text{odd},k \in \text{even}}-\frac{i}{2}[h_j,h_k]
\end{equation}
Note that $h_j$ and $h_k$ commute unless $k = j-1$ or $k = j+1$ which yields
\begin{equation}
     [A,B] = -\frac{i}{2}\sum_{j\in \text{odd}}([h_j,h_{j-1}] + [h_j,h_{j+1}]).
\end{equation}
This can be rewritten as 
\begin{equation}
\begin{aligned}
\relax     [A,B] = -\frac{i}{2}\sum_{j\in \text{odd}}[h_j,h_{j+1}] + \frac{i}{2} \sum_{j \in \text{even}}[h_j,h_{j+1}] =\\  -\frac{i}{2}\sum_j(-1)^{i+1}[h_j,h_{j+1}]
\end{aligned}
\end{equation}
which shows that the first order correction $[A,B]$ yields the same term on even and odd sites, but with an alternating sign. Let's now compute the matrix form of $[h_j,h_{j+1}]$. One readily obtains that the only non-zero matrix elements resulting from this computation are the following transitions
\begin{equation}
\begin{aligned}
    \ket{1011} \rightarrow -\frac{\pi^2}{4}\ket{1101}\\
    \ket{1101} \rightarrow \frac{\pi^2}{4}\ket{1011}\\
\end{aligned}
\end{equation}
all other computational basis states are mapped to 0. Recall that the class of possible deformations introduced in \cite{Khemani2019} are
\begin{equation}
\begin{aligned}
 \sum_{j} Z_{j}\quad
\sum_{j} Z_{j} Z_{j+2} \quad
\sum_{j} Z_{j} Z_{j+3}\\
\sum_{j} P_{j-1} X_{j} P_{j+1}\quad
\sum_{j} P_{j-1} Y_{j} P_{j+1},\\ \sum_{j} P_{j-1} X_{j} P_{j+1} Z_{j+2}\quad
\sum_{j} Z_{j-2} P_{j-1} X_{j} P_{j+1}\\
\sum_{j} P_{j-1} Y_{j} P_{j+1} Z_{j+2}\quad
\sum_{j} Z_{j-2} P_{j-1} Y_{j} P_{j+1}\\ 
\sum_{j} P_{j-1} S_{j}^{+} S_{j+1}^{-} P_{j+2}\quad
\sum_{j} P_{j-1} S_{j}^{-} S_{j+1}^{+} P_{j+2}
\end{aligned}
\end{equation}
The above commutator can be written as 
\begin{equation}
    [h_{j},h_{j+1}] = -\frac{\pi^2}{4}(P_{j} S_{j+1}^{+} S_{j+2}^{-} P_{j+3} - h.c)
\end{equation}
which yields for the first order BCH term

\begin{equation}
\begin{aligned}
     C_1 = \frac{i\pi^2}{8}\sum_j (-1)^{j + 1}(P_{j} S_{j+1}^{+} S_{j+2}^{-} P_{j+3} - h.c)  
\end{aligned}
\end{equation}
Note that the above makes it explicit that the first commutator vanishes when acting on the orbit states and is a consequence of the first order rules being respected in PXP.
Next, consider higher order terms in the expansion that act non-trivially on only 4 qubits. Using the notation $\alpha_j = -\frac{\pi^2}{4}(P_{j} S_{j+1}^{+} S_{j+2}^{-} P_{j+3} - P_{j} S_{j+1}^{-} S_{j+2}^{+} P_{j+3})$ one can write the second order term as
\begin{equation}
    C_2 = -\frac{1}{12}([A,\sum_j(-1)^{j+1}\alpha_j] - [B,\sum_j(-1)^{j+1}\alpha_j]) 
\end{equation}
This expression can be recast as 
\begin{equation}
    -\frac{1}{12}[\sum_j(-1)^{j+1}h_j,\sum_k(-1)^{k+1}\alpha_k]
\end{equation}
Focusing only on terms with support on 4 qubits, one obtains the terms
\begin{equation}
    -\frac{1}{12}\sum_j([h_{j},\alpha_j] - [h_{j+1},\alpha_j])
\end{equation}
Computing first $[h_{j},\alpha_j]$ one obtains that the non zero matrix elements produce the transitions
\begin{equation}
\begin{aligned}
\ket{1111} \rightarrow -\frac{\pi^3}{8}\ket{1101} \\
\ket{1101} \rightarrow -\frac{\pi^3}{8}\ket{1111}
\end{aligned}
\end{equation}
and all other matrix elements vanish. This implies that
\begin{equation}
    [h_{j},\alpha_j] = \frac{\pi^3}{16}(-P_{j+1}X_{j+2}P_{j+3} +Z_{j}P_{j+1}X_{j+2}P_{j+3})
\end{equation}
The second term gives
\begin{equation}
    [h_{j+1},\alpha_j] = \frac{\pi^3}{16}(P_{j}X_{j+1}P_{j+2} -P_{j}X_{j+1}P_{j+2}Z_{j+3})
\end{equation}
By combining the results, one finds that the terms with support on 4 qubits for the second order term in the BCH expansion are 
\begin{equation}
\begin{aligned}
    -\frac{\pi^3}{192}\sum_i (-P_{j+1}X_{j+2}P_{j+3} +P_{j}X_{j+1}P_{j+2}Z_{j+3}\\
    -P_{j}X_{j+1}P_{j+2} +Z_{j}P_{j+1}X_{j+2}P_{j+3})
\end{aligned}
\end{equation}
One can complete the above calculation by also computing terms that will have support on 5 qubits which are given by
\begin{equation}
    -\frac{1}{12}\sum_j(-[h_{j-1},\alpha_j] + [h_{j+2},\alpha_{j}])
\end{equation}
It can be seen that first the term $[h_{j-1},\alpha_j]$ produces the following transitions 
\begin{equation}
\begin{aligned}
\ket{10101}\rightarrow \frac{\pi^3}{8}\ket{11011}\\
\ket{11011} \rightarrow \frac{\pi^3}{8}\ket{10101}
\end{aligned}
\end{equation}
The other term $[h_{j+2},\alpha_j]$ yields the same transitions, but with an added minus sign on both transition which yields for the terms with support on 5 qubits
\begin{equation}
      \frac{\pi^3}{8}\frac{1}{6}\sum_j(P_{j}S_{j+1}^{+}S_{j+2}^{-}S_{j+3}^+P_{j+4} + h.c)
\end{equation}
Up to second order, one thus obtains for the BCH expansion
\begin{equation}
\begin{aligned}
C_0 + C_1 + C_2 = (-\frac{\pi}{2} +\frac{\pi^3}{96})\sum_jP_{j}X_{j+1}P_{j+2} \\
+ i\frac{\pi^2}{8}\sum_j (-1)^{j + 1}(P_{j} S_{j+1}^{+} S_{j+2}^{-} P_{j+3} - P_{j} S_{j+1}^{-} S_{j+2}^{+} P_{j+3}) -\\
\frac{\pi^3}{192}\sum_j ( P_{j}X_{j+1}P_{j+2}Z_{j+3}+Z_{j}P_{j+1}X_{j+2}P_{j+3}) + \\
\frac{\pi^3}{48}\sum_j(P_jS_{j+1}^+S_{j+2}^-S_{j+3}^+P_{j+4} + P_jS_{j+1}^-S_{j+2}^+S_{j+3}^-P_{j+4})
\end{aligned}
\end{equation}

\subsection{Classification of the BCH terms}
The BCH terms obtained from the PXP model can be classified according to which symmetries they respect. The PXP model has three important symmetries which are inversion symmetry about the central bound, time reversal symmetry and a particle-hole like symmetry due to anti-commutation with the operator $\mathcal{P} = \prod_iZ_i$.\\\\
The first order BCH term $[A,B]$ yields 
$\frac{i}{2}\frac{\pi^2}{4}\sum_j (-1)^{j + 1}(P_{j} S_{j+1}^{+} S_{j+2}^{-} P_{j+3} - \text{h.c.})$ which vanishes on the orbit subspace. This term breaks inversion symmetry, time reversal symmetry and doesn't anti-commutes with  $\mathcal{P}$. \\\\
The second order BCH terms is composed of two terms. The first one with support on 4 qubits takes the form $-\frac{1}{24}\frac{\pi^3}{8}\sum_j ( P_{j}X_{j
+1}P_{j+2}Z_{j+3}+Z_{j}P_{j+1}X_{j+2}P_{j+3})$. Such a term respects all symmetries and was shown to improve revivals~\cite{Choi2019} and integrability~\cite{Khemani2019}. The second term with support on 5 qubits is given by  $\frac{1}{6}\frac{\pi^3}{8}\sum_i(P_{j}S_{j+1}^{+}S_{j+2}^{-}S_{j+3}^+P_{j+4} + \text{h.c.})$. This term also respects all symmetries. Note that both these terms act non-trivially on the orbit subspace.
\section{Closing condition for \texorpdfstring{$h_{0,j}$}{}}\label{app:H0Closing}
Provided the fact that there exist an integer $n$ such that $U_{0,j}^n = \mathbb{1}$ and given the decomposition of $h_{0,j}$ in terms of powers of $U_{0,j}$
\begin{equation}
    h_{0,j} = \sum_{k=0}^{n-1}c_kU_{0,j}^k
\end{equation}
it is natural to ask if the local Hamiltonian's $h_{0,j}$ satisfy a closing relation similar to the $U_{0,j}$ closing relation. More precisely, does there exist an integer $m$ such that 
\begin{equation}
    h_{0,j}^m = \sum_{k=0}^{m-1}\alpha_kh_{0,j}^k
\end{equation}
which would restrict the total number of rules of type II one needs to satisfy in order to obtain QMBS phenomenology. First, consider the decomposition of $h_{0,j}^m$ in terms of powers of $U_{0,j}^k$ 
\begin{equation}
    h_{0,j}^s = \sum_{k=0}c_{k}^{(s)}U_{0,j}^k
\end{equation}
where $c_{k}^{(s)}$ denotes the coefficients associated with the $s^{\text{th}}$ power of $h_{0,j}$ and $h_{0,j}^0 \equiv I$. It is straightforward to see that the coefficients $c_{k}^{(s)}$ for $1 \leq s$ are given explicitly by 
\begin{equation}
M^{s-1}
\vec{c}= 
\begin{pmatrix}
c_0^{(s)}\\
c_1^{(s)}\\
\vdots\\
c_{n-1}^{(s)}
\end{pmatrix}
\end{equation}
where 
\begin{equation}
    M = \begin{pmatrix}
c_0 & c_{n-1} & \ldots & c_2 & c_1\\
c_1 & c_0 & c_{n-1} & \ldots & c_2\\
\vdots &   & \ddots &   & \vdots \\
c_{n-1} & c_{n-2} & c_{n-3} & \ldots & c_0
\end{pmatrix}\quad \vec{c} = \begin{pmatrix}
c_0\\
c_1\\
\vdots\\
c_{n-1}
\end{pmatrix} 
\end{equation}
It is a known fact that for any matrix $M$ of size $l$ by $l$, then one has that $M^l$ can always be written as a linear superposition of smaller powers of the matrix $M$. This has the important implication that there exist a set of coefficients $\alpha_k$ such that 
\begin{equation}
h_{0,j}^n = M^n\vec{c} = \sum_{k=0}^{n-1}\alpha_kM^{k}\vec{c} = \sum_{k=0}^{n-1}\alpha_kh_{0,j}^k
\end{equation}
which shows that the local Hamiltonian $h_{0,j}$ closes on itself once the power $n$ is reached.
\section{Spin representation of the models}
The spin representation of the model QMBS-A/B/C and the PXP model is presented in this section using the convention 
\begin{equation}
\begin{aligned}
Z\ket{1} = -\ket{1} \quad
Z\ket{0} = \ket{0}\\
S^+\ket{0} = \ket{1}\quad
S^-\ket{1} = \ket{0}
\end{aligned}
\end{equation}
One has $X = 2S_x$, $Y = 2S_y$ and $Z = 2S_z$ where $S_j$ are the standard spin operators acting on a spin 1/2 particle.

\begin{table}
\begin{center}
\begin{tabular}{||p{2cm}| p{6.4cm} ||}
\hline\hline
QMBS-A &  \\ [0.75ex] 
\hline\hline
Permutation & $\begin{aligned}((3, 13, 11, 7, 9, 5), \\(4, 14, 12, 8, 10, 6))\end{aligned}$ \\
\hline
Phase & (1,1,1,1,1,1,1,1,1,1,1,1,1,1,1,1) \\
\hline
$h_{0,j}$ decomposition & $\begin{aligned}(\frac{\pi}{6} + i\frac{\pi}{2\sqrt{3}})U_{0,j} + (-\frac{\pi}{6} - i\frac{\pi}{6\sqrt{3}})U_{0,j}^2 +\\ \frac{\pi}{12}U_{0,j}^{3} - \frac{\pi}{12}U_{0,j}^0 + \text{h.c}\end{aligned}$  \\
\hline
$U_{0,j}$ spin representation & $\begin{aligned} (S_j^+S_{j+1}^+S_{j+2}^- + S_j^+S_{j+1}^-S_{j+2}^-)\\ +\left( \frac{(I - Z_j)}{2}S_{j+1}^-S_{j+2}^+ + S_j^-S_{j+1}^+\frac{(I - Z_{j+2})}{2}\right)\\ + (S_j^-S_{j+1}^+\frac{(I + Z_{j+2})}{2} + \frac{(I + Z_j)}{2}S_{j+1}^-S_{j+2}^+)\\ + (S_j^+S_{j+1}^+S_{j+2}^+ + S_j^-S_{j+1}^-S_{j+2}^- )^2\end{aligned}$\\
\hline
\hline
QMBS-B &  \\ 
\hline\hline
 Permuation & $\begin{aligned}((1, 15), (2, 16), (3, 9, 5), \\(4, 10, 6), (7, 13, 11),(8, 14, 12)) \end{aligned}$\\ 
 \hline
 Phase & (1,1,1,1,1,1,1,1,1,1,1,1,1,1,1,1)  \\
 \hline
 $h_{0,j}$ decomposition & $\begin{aligned}(\frac{\pi}{6} + i\frac{\pi}{2\sqrt{3}})U_{0,j} + (-\frac{\pi}{6} - i\frac{\pi}{6\sqrt{3}})U_{0,j}^2\\ + \frac{\pi}{12}U_{0,j}^{3} - \frac{\pi}{12}U_{0,j}^0 + \text{h.c}\end{aligned}$  \\
 \hline
 $h_{0,j}$ spin representation & $\begin{aligned}
 \frac{\pi}{4}(S_j^{+}S_{j+1}^{+}S_{j+2}^{+} + S_{j}^{-}S_{j+1}^{-}S_{j+2}^{-}) +\\i\frac{4\pi}{6\sqrt{3}}(S_{j}^{-}S_{j+1}^{+} + S_{j+1}^{-}S_{j+2}^{+} + S_{j+2}^{-}S_{j}^{+})\\ - \frac{\pi P_j}{4} + \text{h.c}\\
 P_j = (S_j^{+}S_{j+1}^{+}S_{j+2}^{+} + S_{j}^{-}S_{j+1}^{-}S_{j+2}^{-})^2
 \end{aligned}$ \\
 \hline
\hline
QMBS-C & \\ 
\hline\hline
Permuation & $\begin{aligned}((3, 5), (4, 6), (7, 15, 9),\\ (8, 16, 10), (11, 13), (12, 14))\end{aligned}$\\ 
\hline
Phase & (1,1,1,1,1,1,1,1,1,1,1,1,1,1,1,1)  \\
\hline
$h_{0,j}$ decomposition & $\begin{aligned}(\frac{\pi}{6} + i\frac{\pi}{2\sqrt{3}})U_{0,j} + (-\frac{\pi}{6} - i\frac{\pi}{6\sqrt{3}})U_{0,j}^2 +\\ \frac{\pi}{12}U_{0,j}^{3} - \frac{\pi}{12}U_{0,j}^0 +\text{h.c}\end{aligned}$  \\
\hline
$h_{0,j}$ spin representation & 
$
\begin{aligned}
\frac{\pi}{2}P_{j+1}X_{j+1}X_{j+2}P_{j+1} + \\(I - P_{j+1})H_{\text{ext},j}(I - P_{j+1}) -\frac{\pi}{2}I,\\
H_{\text{ext},j} = i\frac{4\pi}{6\sqrt{3}}(K_{j+1} + (I-K_{j+1})X_j)\\(K_j + (I - K_j)X_{j+1}X_{j+2}) + \frac{\pi}{4}I + \text{h.c}\\ 
P_{j+1} = (I - Z_{j+1}Z_{j+2})/2,\quad K_j = (I + Z_j)/2
\end{aligned} 
$\\
\hline
\hline
 PXP &  \\ [0.5ex] 
 \hline\hline
 Permuation & $((11, 15), (12, 16))$ \\ 
 \hline
 Phase & (1,1,1,1,1,1,1,1,1,1,i,i,1,1,i,i)  \\
 \hline
 $h_{0,j}$ decomposition & $(\frac{\pi}{4} + i\frac{\pi}{4})U_{0,j} -\frac{\pi}{8}U_{0,j}^2 -  \frac{\pi}{8}I +$ h.c \\
 \hline
 $h_{0,j}$ spin representation & $ -\frac{\pi}{2}P_jX_{j+1}P_{j+2}, \quad P_j = (I-Z_j)/2$ \\
 \hline
 \hline
\end{tabular}
\end{center}
\caption{Spin representation of the models}
\label{tab:PXPTabSpin}
\end{table}

\newpage

\bibliographystyle{apsrev4-1}
\bibliography{automatonscars}
\end{document}